\documentclass[traditabstract]{aa} 
\usepackage[toc,page]{appendix}
\usepackage{graphicx}
\usepackage{natbib}
\bibpunct{(}{)}{;}{a}{,}{,}
\usepackage{txfonts}
\begin{document}
   \title{Gas phase Elemental abundances in Molecular cloudS (GEMS)}
   \subtitle{ I. The prototypical dark cloud TMC~1}

   \author{A.~Fuente\inst{1}
     \and
     D. G.~Navarro\inst{1}
     \and
     P.~Caselli\inst{2}
     \and
     M.~Gerin\inst{3}
     \and
     C.~Kramer\inst{4}
     \and
    E.~Roueff\inst{5}
    \and
%
   T.~ Alonso-Albi \inst{1}
    \and         
    R.~ Bachiller \inst{1}
   \and      
   S.~Cazaux \inst{6}
   \and      
   B.~Commercon\inst{7}
   \and
   R.~Friesen\inst{8}
   \and
   S.~Garc\'{\i}a-Burillo\inst{1}
   \and
   B.~M.~Giuliano\inst{2}
   \and   
   J.~ R.~Goicoechea \inst{9}
   \and
   P. ~Gratier\inst{10}
    \and
    A.~Hacar \inst{11}
    \and
    I.~Jim\'enez-Serra\inst{12}
     \and
    J.~Kirk\inst{13}
    \and
    V.~Lattanzi\inst{2}
     \and
    J.~C.~Loison\inst{14}
     \and
    J.~Malinen\inst{15,16}
     \and
     N.~Marcelino\inst{9}
     \and
     R.~ Mart\'{\i}n-Dom\'enech\inst{17}
     \and
     G.~Mu\~noz-Caro\inst{12}
     \and
    J.~Pineda\inst{2}
    \and
    M.~Tafalla\inst{1}
    \and
    B.~Tercero\inst{1}
    \and
    D.~Ward-Thompson\inst{18}
    \and
    S.~P. Trevi\~no-Morales\inst{19}
     \and
    P.~Rivi\'ere-Marichalar\inst{1}
     \and
     O.~Roncero\inst{9}
    \and
    T.~Vidal\inst{10}
    \and
    Maikel Y. Ballester\inst{20}
    }
              
   \institute{ 
Observatorio Astron\'omico Nacional (OAN), Alfonso XII, 3,  28014, Madrid, Spain
    \and
Centre for Astrochemical Studies, Max-Planck-Institute for Extraterrestrial Physics, Giessenbachstrasse 1, 85748, Garching, Germany    
\and
Observatoire de Paris, PSL Research University, CNRS, \'Ecole Normale Sup\'erieure, Sorbonne Universit\'es, UPMC Univ. Paris 06, 75005, Paris, France
\and
Instituto Radioastronom\'{\i}a Milim\'etrica (IRAM), Av. Divina Pastora 7, Nucleo Central, 18012, Granada, Spain
\and
Sorbonne Universit\'e, Observatoire de Paris, Universit\'e PSL, CNRS, LERMA, F-92190,  Meudon, France
\and
Faculty of Aerospace Engineering, Delft University of Technology, Delft, The Netherlands ; University of Leiden, P.O. Box 9513, NL, 2300 RA, Leiden, The Netherlands 
\and
\'Ecole Normale Sup\'erieure de Lyon, CRAL, UMR CNRS 5574, Universit\'e Lyon I, 46 All\'ee d'Italie, 69364, Lyon Cedex 07, France 
\and
National Radio Astronomy Observatory, 520 Edgemont Rd., Charlottesville VA USA 22901
\and
Instituto de F\'{\i}sica Fundamental (CSIC), Calle Serrano 123, 28006, Madrid, Spain
\and
Laboratoire d'astrophysique de Bordeaux, Univ. Bordeaux, CNRS, B18N, all\'ee Geoffroy Saint-Hilaire, 33615, Pessac, France  
\and
Leiden Observatory, Leiden University, PO Box 9513, 2300-RA, Leiden, The Netherlands 
\and
Centro de Astrobiolog\'{\i}a (CSIC-INTA), Ctra. de Ajalvir, km 4, Torrej\'on de Ardoz, 28850, Madrid, Spain
\and
Department of Physics, University of Warwick, Coventry CV4 7AL, UK
\and
Institut des Sciences Mol\'eculaires (ISM), CNRS, Univ. Bordeaux, 351 cours de la Lib\'eration, F-33400, Talence, France
\and
Department of Physics, University of Helsinki, PO Box 64, 00014, Helsinki, Finland
\and
Institute of Physics I, University of Cologne, Cologne, Germany
\and
Harvard-Smithsonian Center for Astrophysics, Cambridge, MA 02138, USA
\and
Jeremiah Horrocks Institute, University of Central Lancashire, Preston PR1 2HE, UK 
\and
Chalmers University of Technology, Department of Space, Earth and Environment, SE-412 93 Gothenburg, Sweden
\and 
Departamento de F{\'i}sica, Universidade Federal de Juiz de Fora-UFJF, Juiz de Fora,
MG 36036-330, Brazil.
}
 
 \abstract  {GEMS is an IRAM 30m Large Program whose aim is determining the elemental depletions and the ionization fraction
 in a set of prototypical star-forming regions. This paper presents the first results from the 
prototypical dark cloud TMC~1. Extensive millimeter observations have been carried out with the IRAM 30m telescope (3\,mm and 2\,mm) and the 40m Yebes  telescope (1.3\,cm and 7\,mm) to determine the fractional abundances of CO, HCO$^+$, HCN, CS, SO, HCS$^+$, and N$_2$H$^+$  in three cuts which intersect the dense filament at the well-known positions
TMC~1-CP, TMC~1-NH3, and TMC~1-C, covering a visual extinction  range from A$_V$~$\sim$~3 to $\sim$20 mag. Two phases with differentiated chemistry can be distinguished: i)  the translucent envelope with molecular hydrogen densities of 1$-$5$\times$10$^{3}$~cm$^{-3}$;   and ii) the dense phase, located at A$_V >$~10 mag, with molecular hydrogen densities $>$10$^4$~cm$^{-3}$. Observations and modeling  show that the gas phase abundances of C and O progressively decrease along the  C$^+$/C/CO transition zone (A$_V \sim$~3~mag) where C/H~$\sim$~8$\times$10$^{-5}$ and C/O$\sim$0.8$-$1, until the beginning of the dense phase at A$_V$ $\sim$ 10 mag. This is consistent with the grain temperatures being below the CO evaporation temperature in this region. In the case of sulfur, a strong depletion should occur before the translucent phase where we estimate a  S/H ~$\sim$~(0.4 - 2.2) $\times$10$^{-6}$, an abundance $\sim$7-40 times lower than the solar value. A second strong depletion must be present during the formation of the thick icy mantles to achieve the values of S/H measured in the dense cold cores (S/H~$\sim$8$\times$10$^{-8}$).  Based on our chemical modeling, we constrain the value of  $\zeta_{\rm H_2}$ to $\sim$ (0.5 - 1.8) $\times$10$^{-16}$ s$^{-1}$ in the translucent cloud.}
  
   \maketitle
%

\section{Introduction}
In recent years, space telescopes such as Spitzer and Herschel have revolutionized 
our view of star-forming regions. Images of giant molecular clouds 
and dark cloud complexes have revealed  spectacular networks of filamentary structures where stars 
are born \citep{Andre2010}.
Interstellar filaments are almost everywhere in the Milky Way and are the preferred site for star formation. Now we believe that 
filaments precede the onset of most star formation, funneling interstellar gas and dust into 
increasingly denser concentrations that will contract and fragment, leading to gravitationally bound 
prestellar cores that will eventually  form stars.

Gas chemistry plays a key role in the star formation process 
by regulating fundamental parameters such as the gas cooling rate, and the gas ionization fraction. Molecular filaments can
fragment into prestellar cores to a large extent because molecules cool the gas, thus diminishing the thermal
support relative to self-gravity. The ionization fraction controls the coupling of magnetic fields 
with the gas, driving the dissipation of turbulence and angular momentum transfer, and therefore it 
plays a crucial role in the cloud collapse (isolated vs clustered star formation) and the dynamics 
of accretion disks (see \citealp{Zhao2016, Padovani2013}). 
In the absence of other ionization agents (X-rays, UV photons, J-type shocks), 
the steady-state ionization fraction is proportional to $\sqrt{\zeta _{H_2}/n}$, where $n$ is the molecular hydrogen density and 
$\zeta _{H_2}$ is the cosmic-ray ionization rate for H$_2$ molecules, which becomes an essential parameter in molecular 
cloud evolution \citep{Oppenheimer1974, Kee1989, Caselli2002}.
The gas ionization fraction, X(e$^-$)=n(e$^-$)/n$_{\rm H}$, and the molecular abundances depend on the elemental 
depletion factors \citep{Caselli1998}. In particular, carbon (C) is the main donor of electrons 
in the cloud surface (A$_v$$<$4 mag) and, because of its lower ionization potential and as long as 
it is not heavily depleted, sulfur (S) is the main donor in the $\sim$3.7$-$7 magnitude range that 
encompasses a large fraction of the molecular cloud mass \citep{Goicoechea2006}. 
Since CO and CII are the main coolants, depletions of C 
and O determine the gas cooling rate in molecular clouds.
Elemental depletions also constitute
a valuable piece of information for our understanding of the grain composition and evolution. For a given element X,
the missing atoms in gas phase are presumed to be locked up in solids, i.e., dust grains and/or icy mantles. The
knowledge of the elemental depletions would hence provide a valuable
information to study the changes in the dust grain composition across the cloud.
Surface chemistry and the interchange of molecules between the solid and gas phases have a leading role
in the gas chemical evolution from the diffuse cloud to the prestellar core phase.
 
The Gas phase Elemental abundances in Molecular
cloudS (GEMS) is an IRAM 30m Large Program whose
aim is estimating the S, C, N, O depletions and X(e$^-$)
as a function of visual extinction, in a selected set of prototypical star-forming filaments. Regions 
with different illumination are included in the sample in order to investigate the influence 
of UV radiation (photodissociation, ionization, photodesorption) and turbulence (grain sputtering,
grain-grain collisions) on these parameters, and eventually
in the star formation history of the cloud. This is the first of a series of GEMS papers and
it is dedicated to the prototypical dark cloud TMC 1.

\section{TMC~1}
The Taurus molecular cloud (TMC), at a distance of 140 pc \citep{Elias1978,
Onishi2002}, is one of the closest molecular cloud complexes, and is considered 
an archetype of low-mass star-forming regions. It has been the target of several cloud evolution
and star formation studies \citep{Unge1987, Mizuno1995, Goldsmith2008}, being
extensively mapped in CO \citep{Cernicharo1987, Onishi1996, Narayanan2008} 
and visual extinction \citep{Cam1999, Padoan2002}. The most massive
molecular cloud in Taurus is the Heiles cloud 2 (HCL 2) \citep{Onishi1996}.
TMC~1 was included  in the Herschel Gould Belt Survey \citep{Andre2010}.
One first analysis of these data were carried out by \citet{Malinen2012} 
who generated visual extinction maps of the two long filaments in HCL 2
based on near-IR (NIR) extinction and Herschel data. As one of the most
extensively studied molecular filament, TMC~1 is also included
in the Green-Bank Ammonia Survey (PIs: R. Friesen \& J. Pineda) \citep{Friesen2017}.

TMC~1 has been also the target of numerous chemical studies. In particular,
the positions TMC~1-CP and TMC~1-NH3  (the cyanopolyne and ammonia
emission peaks) are generally adopted as templates to compare with chemical 
codes \citep{Feher2016, Gratier2016, Agundez2013}. Less studied from
the chemical point of view, TMC~1-C has been identified as an accreting
starless core \citep{Schnee2007, Schnee2010}.

\begin{figure}
\includegraphics[angle=0,scale=.6]{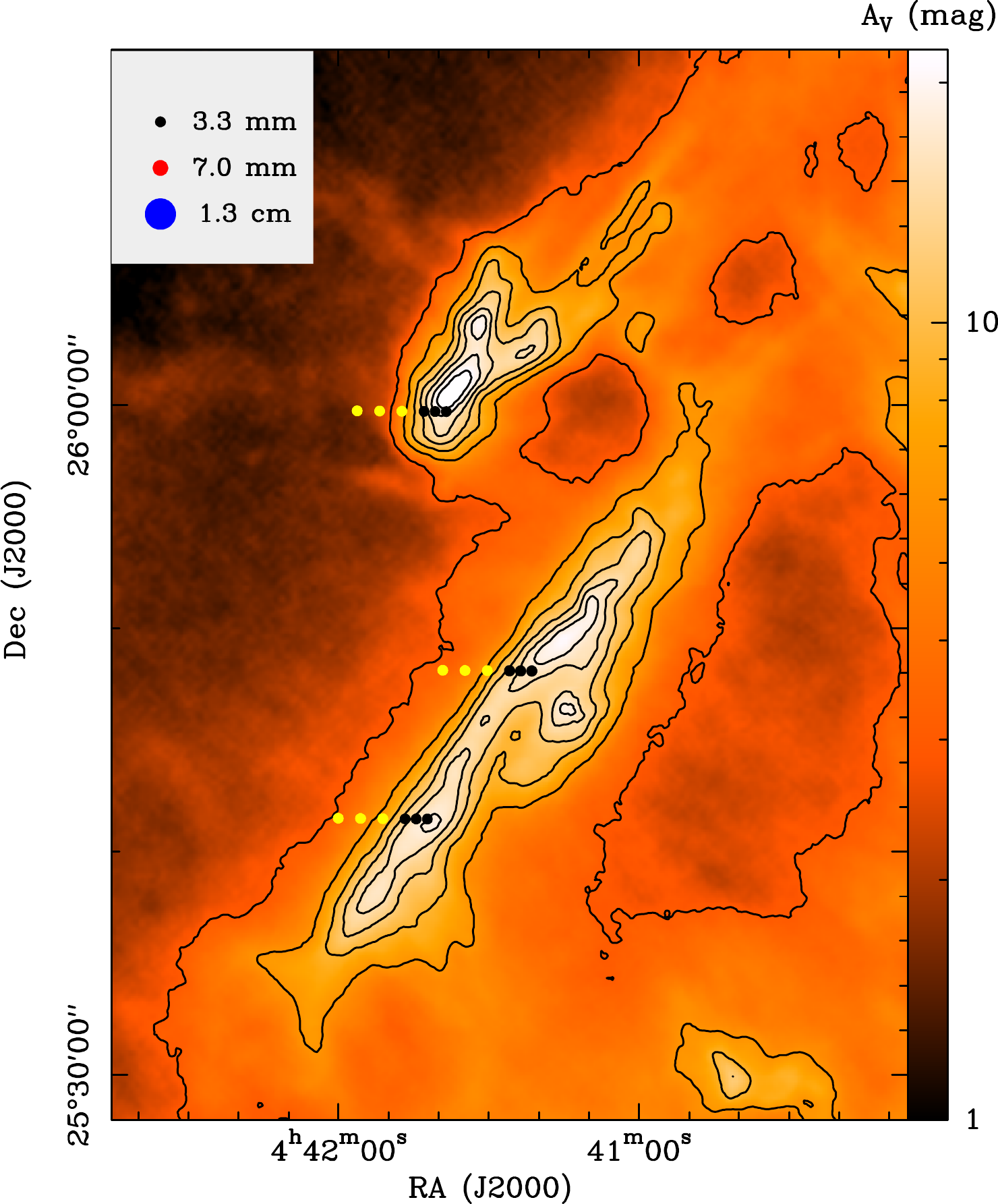}
\caption{ Visual extinction map of TMC 1 (Kirk et al., in prep). The positions observed with the 30m telescope are
indicated with circles. Black circles mark positions observed only with the 30m telescope while yellow circles
show the positions observed also with the Yebes 40m telescope. In the upper left corner, we show the beam of the IRAM telescope at 3\,mm (beam$\sim$29$"$) and of Yebes 40m telescope at 1.3\,cm (beam$\sim$84$"$) filled and at 7\,mm (beam$\sim$42$"$). Contours are 3, 6, 9, 12, 15, and 18~mag. }
\label{map}
\end{figure}

\begin{table}
\begin{centering}
\begin{tabular}{l|l|l}
\multicolumn{3}{l}{{\small Table 1.- Molecular tracers used in this study}} \\ \hline 
\multicolumn{1}{c|}{} & \multicolumn{1}{c|}{ A$_V <$ 10 mag} & \multicolumn{1}{c}{ A$_ V >$ 10 mag}    \\ \hline \hline 
 X(e$^-$)     &    $^{13}$CO, HCO$^+$, H$^{13}$CO$^+$   &   C$^{18}$O, HC$^{18}$O$^+$, N$_2$H$^+$     \\  
 n(H$_2$)    &   CS, C$^{34}$S                                             &   C$^{34}$S, $^{13}$CS, SO                   \\ 
 C/H             &   $^{13}$CO, HCN, CS                                   &   C$^{18}$O, H$^{13}$CN           \\
 O/H             &   $^{13}$CO, SO                                             &   C$^{18}$O, SO, $^{34}$SO                        \\
 N/H             &    HCN                                                             &    H$^{13}$CN, N$_2$H$^+$               \\ 
 S/H             &    CS, C$^{34}$S, SO, HCS$^+$                    &    C$^{34}$S, $^{13}$CS, SO, $^{34}$SO       \\
\hline 
\end{tabular}
\end{centering}
\end{table}

\section{Observational Strategy}

In order to derive the elemental gas abundance of C, O, N and S, we need to determine the abundances of the
main gas reservoirs (see Table~1). 
Essentially, most of the carbon in molecular clouds is locked in CO and the C depletion
is derived from the study of CO and its isotopologues. Several works have studied the 
depletion of CO in dense starless cores and young protostars 
\citep{Caselli1999,Kramer1999,Bacmann2002,Alonso2010,Hernandez2011,Maret2013,Miet2013,Lippok2013}.
The main reservoirs of nitrogen are atomic nitrogen (N) and molecular nitrogen (N$_2$) which are not observable. 
The nitrogen abundance needs to be derived by applying a chemical model to fit the observed abundances
of nitriles (HCN, HNC, CN) and N$_2$H$^+$. The HCN abundance is also dependent on 
the amount of atomic C in gas phase and hence, on the C/O ratio \citep{Loison2014}. 
Abundant oxygenated species such as O, O$_2$, H$_2$O and OH, cannot be observed in the 
millimeter domain  and the oxygen depletion should be indirectly derived 
through comparison with chemical models, as well. In the case of sulfur, depending 
on the local physical conditions and the chemical age,
atomic S and/or SO are expected to be the main gas phase reservoir in
dense clouds \citep{Fuente2016, Vidal2017}. Unfortunately, the direct observation of atomic S is
difficult and, thus far, has only been detected in 
some bipolar outflows using the infrared space telescope Spitzer \citep{Anderson2013}. Sulfur can also 
be traced by sulfur recombination lines but they are very weak and not easy to interpret \citep{Roshi2014}.
Sulfur depletion in molecular clouds is determined from the observation of a few 
molecular compounds, mainly CS, HCS$^+$ or SO (see, e.g., \citealp{Goicoechea2006}), whose abundances are 
very sensitive to the C/O gas-phase ratio and also time evolution. Its determination hence
requires a good characterization of the gas physical and chemical conditions. 

For the present study, we selected a subset of species (CO, HCO$^+$, HCN, CS,
SO, HCS$^+$, and N$_2$H$^+$) that are essential to derive the elemental abundances 
in the molecular gas.
The observations were performed using the receiver setups listed in Table~B.1 
and the observed molecular transitions are shown in Table~B.2. When possible we observe several lines of the same species 
in order to accurately determine the molecular abundance. When only one line was observed,
we use the molecular hydrogen density derived from the fitting of the CS (and their rarer 
isotopologues C$^{34}$S and $^{13}$CS) 3$\rightarrow$2 and 2$\rightarrow$1 lines. Towards 
the edge of the cloud, the densities are lower and the CS 3$\rightarrow$2 line is not detected.
For this reason we complement the 30m observations with the CS 1$\rightarrow$0
line as observed with the 40m Yebes telescope. 
The 40m configuration allows us to observe simultaneously the NH$_3$ (1,1) and (2,2) lines
in band K. We use these observations to constrain the gas kinetic temperature at the cloud edges. 

The high sensitivity required by our project, prohibits the mapping of a large area. 
Instead, we observe the three right-ascension cuts covering visual extinctions 
between A$_V$$\sim$ 3 mag to $\sim$ 20 mag (see  Fig.~\ref{map}). 
In details, we have observed 6 positions per cut which corresponds
to the offsets (0",0"), (+30",0), (+60",0"), (+120",0"), (+180",0") and (+240",0") relative to the positions listed in
Table~2,  
In addition to the 30m observations, we carried out observations with the 40m Yebes telescope
towards the positions marked with yellow circles in Fig.~\ref{map}.

\section{Data acquisition}

\subsection{IRAM 30m telescope}

The 3\,mm and 2\,mm observations were carried out using 
the IRAM 30-m telescope at Pico Veleta (Spain) during 
three observing periods in July 2017, August 2017 and February 2018.  
The telescope parameters at 3\,mm and 2\,mm are listed in Table~B.1
with the beam size varying with the
frequency as  HPBW($"$)=2460 /$\nu$ where $\nu$ is in GHz.
The observing mode was frequency switching with a frequency throw of 6 MHz
well adapted to remove standing waves between the secondary and the receivers. 
The Eight MIxer Receivers (EMIR) and the Fast Fourier Transform Spectrometers (FTS) with a
spectral resolution of 49~kHz were used for these observations. 
The intensity scale is T$_{MB}$ which is a good estimate of T$_{B}$
as long as the source size is comparable to the observational beam. In our case, the
emission is expected to be more extended. In the limiting case of the source being extended
through the whole sky, the correct intensity scale would be  T$_{A}^*$, where 
T$_{MB}$ and T$_A^*$ are related by $T_{MB} =(F_{eff} / B_{eff})T_A^*$ (see Table B.1).
Since the difference between one scale and the other is not large, $\approx$ 17 \% at 86 GHz
and 27\% at 145 GHz, we adopt the T$_{MB}$ scale. The uncertainty in the source size 
is included in the line intensity errors which are assumed to be $\sim$20\%. Although 
numerous lines are detected in the range of frequencies covered by our observations,
in this paper we concentrate on the most
abundant molecules (and their isotopologues): CO, HCO$^+$, HCN, CS, SO,
HCS$^+$, and N$_2$H$^+$. Other species will be analyzed in
forthcoming papers.

\begin{table}
\begin{tabular}{llll}\\
\multicolumn{4}{l}{Table 2. Source coordinates} \\ \hline \hline
\multicolumn{1}{c}{} & \multicolumn{1}{c}{RA(J2000)} & 
\multicolumn{1}{c}{Dec (J2000)}  &  \multicolumn{1}{c}{V$_{lsr}$ (km s$^{-1}$)}
\\ \hline \hline                                                  
TMC~1-CP     &  04$^{\rm h}$41$^{\rm m}$41$^{\rm s}$.90    &  25$^{\circ}$41$'$27$''$.1  &   5.8    \\     
TMC~1-NH3   &  04$^{\rm h}$41$^{\rm m}$21$^{\rm s}$.30    &  25$^{\circ}$48$'$07$''$.0  &   5.8    \\
TMC~1-C       &  04$^{\rm h}$41$^{\rm m}$38$^{\rm s}$.80    &  25$^{\circ}$59$'$42$''$.0  &   5.2    \\
\hline \hline
\end{tabular}
\end{table}

\subsection{Yebes 40m telescope}
The RT40m is equipped with HEMT receivers for 
the 2.2-50 GHz range, and a SIS receiver for the 85-116 GHz range. Single-dish 
observations in K-band (21-25 GHz) and Q-band (41-50 GHz) can be performed
simultaneously. This configuration was used to observe the positions 
marked with a yellow circle in Fig.~1. The backends consisted of FFTS 
covering a bandwidth of $\sim$2 GHz in band K and $\sim$9 GHz in band Q,
with a spectral resolution of $\sim$38 kHz.
Central frequencies were 23000~MHz and 44750~MHz for the 
K and Q band receivers, respectively.
The observing procedure was position-switching, and the OFF-positions are 
RA(J2000) = 04$^{\rm h}$42$^{\rm m}$24$^{\rm s}$.24 Dec(2000):25$^{\circ}$41$'$27$''$.6 for TMC~1-CP,
RA(J2000) = 04$^{\rm h}$42$^{\rm m}$29$^{\rm s}$.52  Dec(2000):25$^{\circ}$48$'$07$''$.2 for TMC~1-NH3,
RA(J2000) = 04$^{\rm h}$42$^{\rm m}$32$^{\rm s}$.16  Dec(J2000):25$^{\circ}$59$'$42$''$.0 for TMC~1-C. 
These positions were checked to be empty of emission before the observations. 
The intensity scale is T$_{MB}$ with conversion factors of 4.1 Jy/K in band K 
(T$_{MB}$/T$_A^*$=1.3) and in 5.7 Jy/K in band Q (T$_{MB}$/T$_A^*$=2.1).
The HPBW of the telescope is 42$"$ at 7\,mm and 84$"$ at 1.3\,cm (Table~B.1). 

\subsection{Herschel Space Observatory: A$_V$ and T$_d$ maps}
In this work, we use  the column density and dust temperature maps of TMC~1 
created following the process described 
in \citet{Kirk2013} and Kirk et al. (in prep). In the following, we give a brief explanation of the methodology.
The PACS and SPIRE \citep{Poglitsch2010, Griffin2010} data were taken 
as part of the Herschel Gould Belt Survey \citep{Andre2010} and were reduced as described 
in Kirk et al. (in prep). The absolute calibration (median flux level) of the maps was 
estimated using data from Planck and IRAS (c.f. \citealp{Bernard2010}). The data was then 
convolved to the resolution of the longest wavelength 500~$\mu$m (36 arcsec).

A modified blackbody function of the form $F_\nu = M B_\nu(T) \kappa_\nu / D^2$ was fitted to each 
point where $M$ is the dust mass, $B_\nu (T)$ is the Planck function at temperature $T$, and 
$D=140$~pc was the assumed distance to Taurus. The dust mass opacity was assumed to follow 
a standard law, $\kappa_\nu \propto \nu^\beta$, with $\beta=2$ and a reference value 
of 0.1 cm$^2$ g$^{-1}$ at $\lambda$=1~THz \citep{Beckwith1990}.
When using the same dust assumptions, the resulting dust map agreed well with the Planck 353~GHz
Optical Depth map above $N(H_2) \sim 1.5\times10^{21}$ cm$^{-2}$ (Kirk et al., in prep). The typical uncertainty 
on the fitted dust temperature was 0.3-0.4 K. The uncertainty on the column density was 
typically 10\% and reflects the assumed calibration error of the Herschel maps (Kirk et al., in prep).

\section{Spectroscopic data: Line profiles}
Figures~\ref{cp-espec1} to ~\ref{c2-espec2} show a subset of our spectra across the cuts TMC~1-CP, TMC~1-NH3 and TMC~1-C. The 
lines of the most abundant species are optically thick  at A$_v$$>$ 7 mag and 
present self-absorbed profiles.  However, only the lines of the main isotopologue are detected
towards positions with  A$_v$$<$ 7 mag.  Linewidths vary between $\sim$0.3 to $\sim$1.5~km~s$^{-1}$ depending on the transition. 
The largest line widths are measured in the $^{13}$CO 1$\rightarrow$0 lines with 
$\Delta$v$\sim$1.5$\pm$0.5 km~s$^{-1}$. 
The higher excitation lines  of species like CS and SO, and those of  the high dipole moment tracers
HCS$^+$, N$_2$H$^+$ and HCN, show $\Delta$v $\sim$0.4$\pm$~0.1~km~s$^{-1}$. 
Similar linewidths are observed in the NH$_3$ (1,1) and (2,2) inversion lines. 
Line broadening because of high optical depths might explain, at least partially, the large linewidths 
observed in the $^{13}$CO and CS lines. High optical depths are also measured in the
HCN 1$\rightarrow$0 lines but the linewidth remains  narrow, $\Delta$v $\sim$0.4$\pm$~0.1~km~s$^{-1}$.
The narrow HCN linewidths are better understood as the consequence of the existence of layers with 
different excitation and hence chemical conditions along the line of sight.

\begin{center}
\begin{figure}
\includegraphics[angle=0,scale=.7]{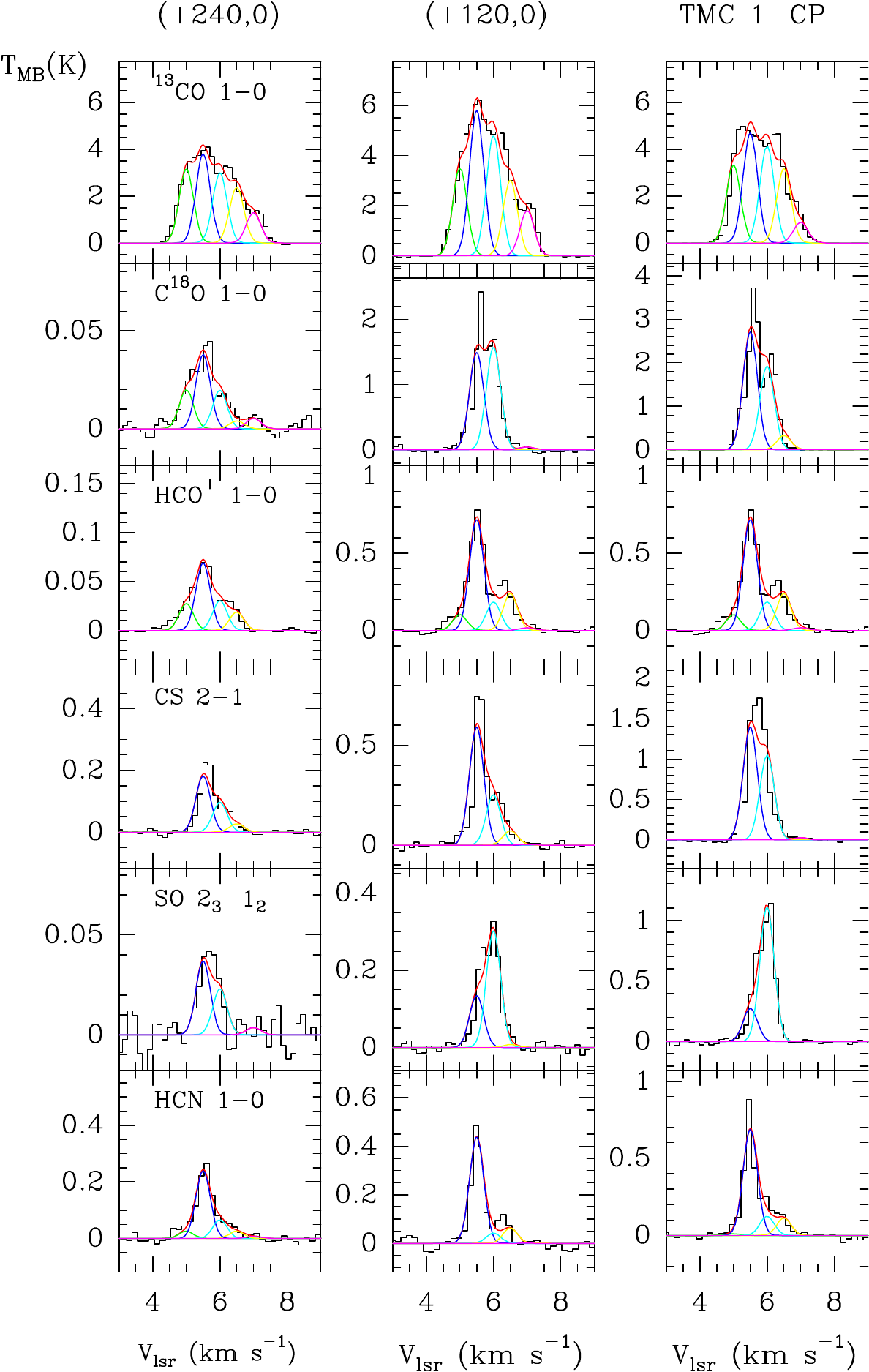}
\caption{Selection of 30m spectra towards the offsets (+240",0), (120",0) and (0,0) in the
TMC~1-CP cut. In order to investigate the velocity structure we have fitted the observed line
profiles with 5 Gaussians with a fixed linewidth of 0.5~km~s$^{-1}$, each centered at the velocities
5.0 (green), 5.5 (dark blue), 6.0 (light blue), 6.5 (yellow), 7.0~km~s$^{-1}$ (fucsia).  } 
\label{profiles}
\end{figure}
\end{center}

Several authors have discussed the  complex velocity structure of the TMC 1 cloud
\citep{Lique2006, Feher2016,Dobashi2018}. Based on high-velocity resolution ($\delta v_{lsr}$=0.0004 km~s$^{-1}$)
observations of the HC$_3$N J=5$\rightarrow$4 line, \citet{Dobashi2018} propose that the dense 
TMC~1 filament is composed of at least 4 velocity 
components at v$_{lsr}$=5.727, 5.901, 6.064  and 6.160 km~s$^{-1}$ with small linewidths, $\sim$0.1 km~s$^{-1}$, 
and a more diffuse component at 6.215 km~s$^{-1}$ with a linewidth of $\sim$0.5 km~s$^{-1}$. 
The velocity resolution of our observations (from 0.27 km s$^{-1}$ at 7mm, to 0.16 km s$^{-1}$ at 3\,mm
and 0.09 km~s$^{-1}$ at 2\,mm ) is not enough to resolve these narrow velocity components. 
In spite of this, in order to have a deeper insight in the velocity structure of the region,
we have fitted the observed profiles using 5 velocity components centered at 
v$_{lsr}$=5.0, 5.5, 6.0, 6.5, 7.0 km s$^{-1}$
and with a fixed $\Delta$ v=0.5 km~s$^{-1}$. Fig.~\ref{profiles} shows the result of our fitting  
for three positions, offsets (+240",0), (+120",0) and (0,0) in the cut across TMC 1-CP. 
Interestingly, the number of velocity components detected in each transition remains constant with
the position, even when the position (+240",0) is located 0.16~pc away from TMC~1-CP.
However, the number of detected velocity components does vary from one
transition (and species) to another. The five velocity components appear in the spectra
of $^{13}$CO 1$\rightarrow$0, C$^{18}$O 1$\rightarrow$0, and HCO$^+$ 1$\rightarrow$0.
The CS 2$\rightarrow$1 and SO 2$_3$$\rightarrow$1$_2$ spectra present intense emission 
in the  5.5 km~s$^{-1}$ and 6.0 km~s$^{-1}$ components and weak wings at the velocity of the other components.
Interestingly, only the 5.5 km~s$^{-1}$ component presents intense emission in the HCN spectra. 
As a first approximation,
in this paper we use the integrated line intensities to derive column densities and abundances. 
This is motivated by the limited spectral resolution (at 3\,mm) and sensitivity (at 2\,mm) of our observations 
that would introduce large uncertainties in the multi-velocity analysis. Taking into account the analysis 
of the line profiles described in this section, we can conclude that our results probe the moderate 
to high-density gas detected
at V$_{lsr}$=5.5 km~s$^{-1}$ and 6.0 km~s$^{-1}$. 

\begin{figure*}
\includegraphics[angle=0,scale=.22]{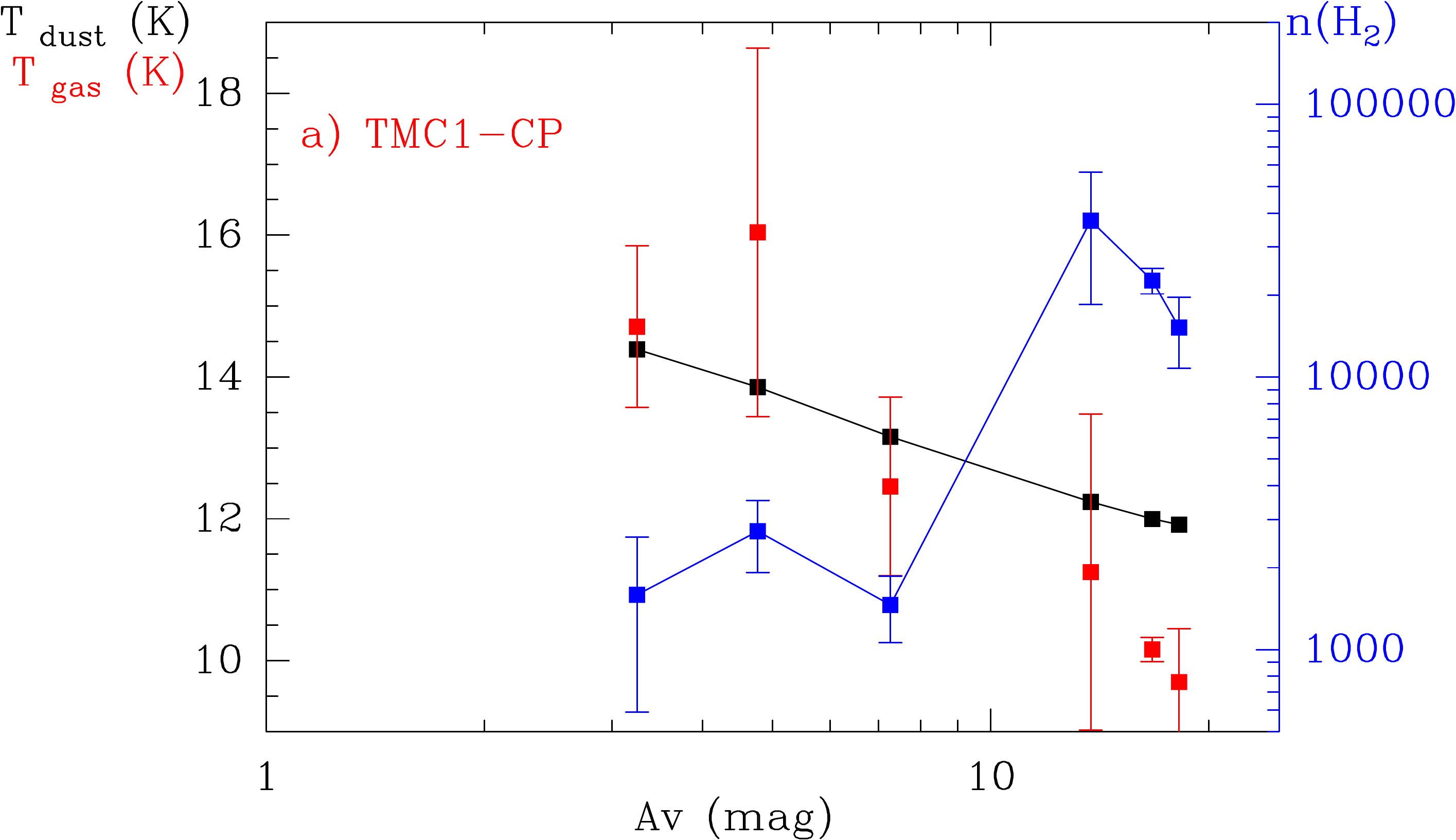}
\includegraphics[angle=0,scale=.22]{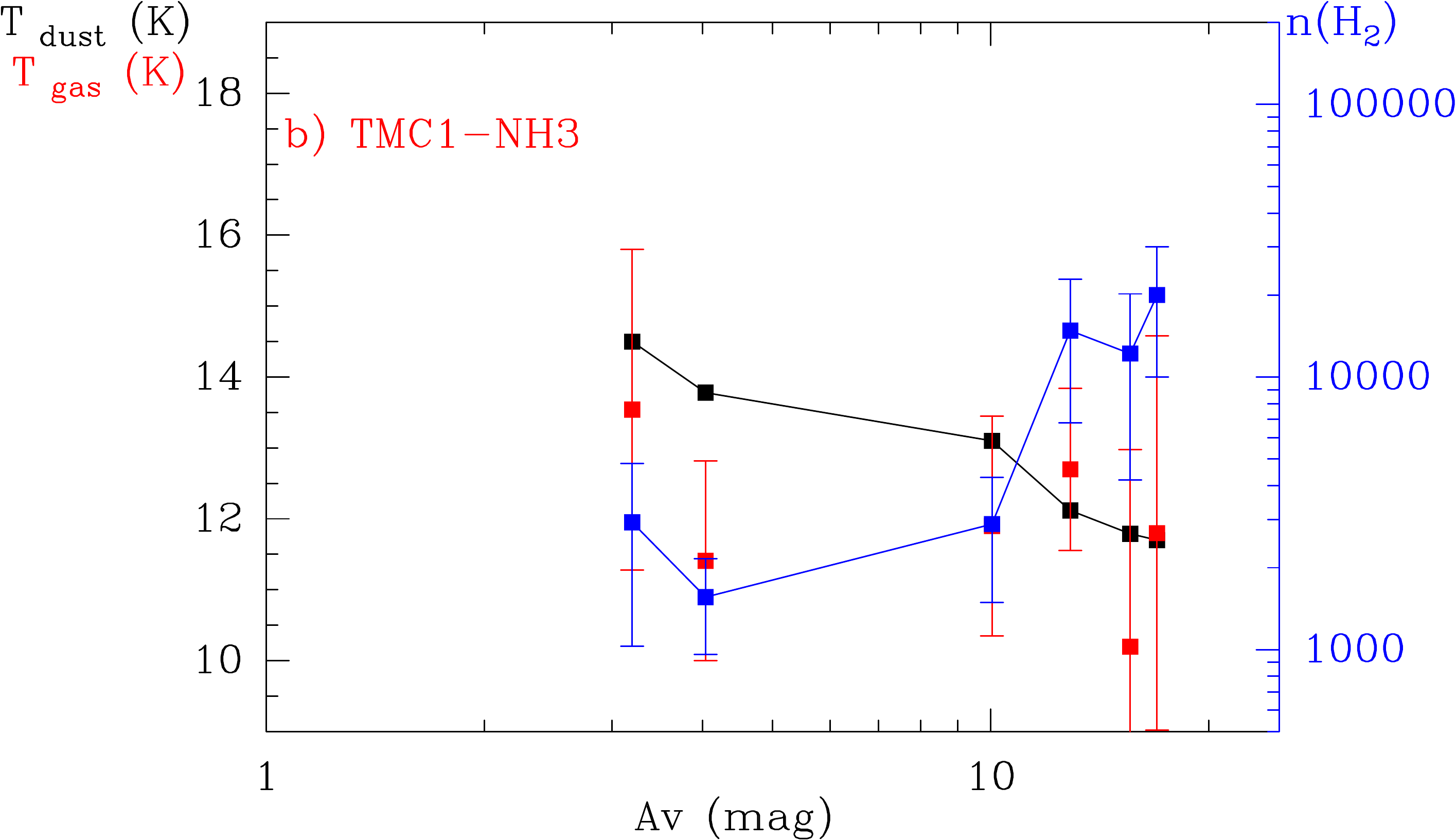} 
\includegraphics[angle=0,scale=.22]{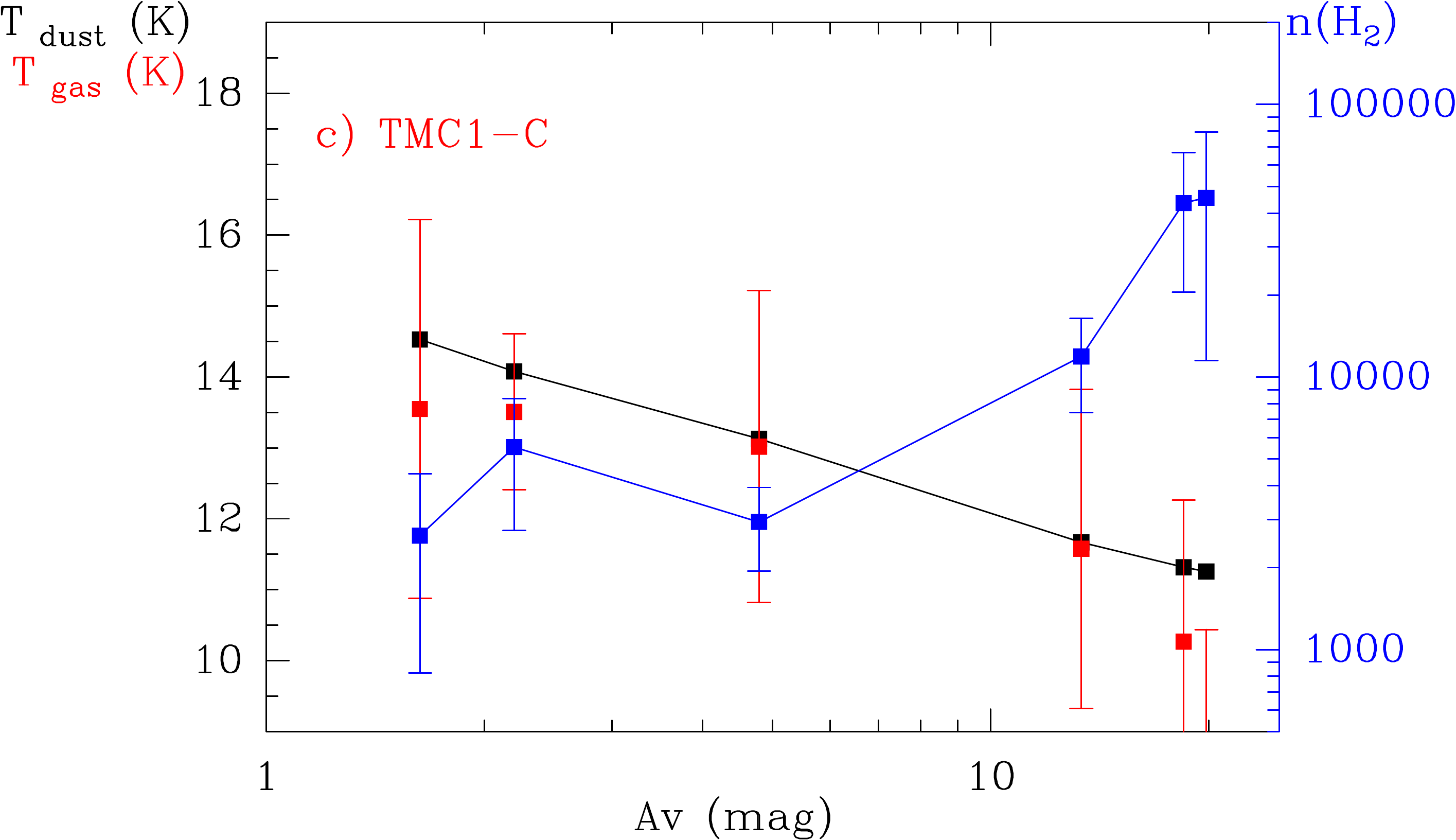}
\caption{Estimated values of the dust temperature (black) as derived from Kirk et al. (in prep),  gas temperature (red) and molecular hydrogen density (blue) across the cuts through TMC~1-CP (a), TMC~1-NH3 (b) y TMC~1-C (c).}
\label{nh2}
\end{figure*}

\begin{table*}
\label{phys}
\begin{tabular}{lll | ll | ll | l}\\
\multicolumn{8}{l}{Table 3. Physical conditions} \\ \hline \hline
\multicolumn{1}{c}{}     & \multicolumn{2}{c|}{HSO (Kirk+19)} &
\multicolumn{2}{c|}{GEMS (This work)} & \multicolumn{2}{c|}{NH$_3$ (Feh\'er+16)}  & 
\multicolumn{1}{c}{Species and transitions used}  \\   
\multicolumn{1}{c}{Position} & \multicolumn{1}{c}{T$_d$} & \multicolumn{1}{c|}{N(H$_2$)} & 
\multicolumn{1}{c}{T$_k$}   & \multicolumn{1}{c|}{n(H$_2$)} & 
\multicolumn{1}{c}{T$_k$}   & \multicolumn{1}{c|}{n(H$_2$)}  &   
\multicolumn{1}{c}{in the calculations}  \\ 
\multicolumn{1}{c}{}     & \multicolumn{1}{c}{(K)} & \multicolumn{1}{c|}{($\times$10$^{21}$ cm$^{-2}$)} & 
\multicolumn{1}{c}{(K)}  & \multicolumn{1}{c|}{($\times$10$^4$ cm$^{-3}$)} &  
\multicolumn{1}{c}{(K)}  & \multicolumn{1}{c|}{($\times$ 10$^4$ cm$^{-3}$)} & 
\multicolumn{1}{c}{} \\    \hline
TMC1-CP+0    &      11.92   &    18.20   &  9.7$\pm$0.8	  &  1.5$\pm$0.4	& 10.6$\pm$1.1 & 1.0$\pm$0.3  &  $^{13}$CS - C$^{34}$S J=2$\rightarrow$1, 3$\rightarrow$2     \\
TMC1-CP+30   &      12.00   &    16.71   & 10.2$\pm$0.2	  &  2.3$\pm$0.3    &     &   &  $^{13}$CS	- C$^{34}$S  J=2$\rightarrow$1, 3$\rightarrow$2   \\
TMC1-CP+60   &      12.24   &    13.74   & 11.2$\pm$2.0	  &  3.7$\pm$1.9    &     &   &  $^{13}$CS	- C$^{34}$S  J=2$\rightarrow$1, 3$\rightarrow$2    \\  
TMC1-CP+120  &      13.16   &    7.27    & 12.5$\pm$1.2	  &  0.15$\pm$0.04  &     &   &  C$^{34}$S - CS    J=1$\rightarrow$0, 2$\rightarrow$1, 3$\rightarrow$2       \\ 
TMC1-CP+180  &      13.86   &    4.77    & 16.0$\pm$3.0	  &  0.27$\pm$0.09  &     &   &  C$^{34}$S - CS   J=1$\rightarrow$0, 2$\rightarrow$1, 3$\rightarrow$2         \\
TMC1-CP+240  &      14.39   &    3.25    & 14.7$\pm$1.1	  &  0.16$\pm$0.10  &     &   &  CS         J=1$\rightarrow$0, 2$\rightarrow$1, 3$\rightarrow$2    \\ \hline
TMC1-NH3+0    &      11.70   &    16.97   & 11.8$\pm$2.8   &  2.0$\pm$1.0   &  11.0$\pm$1.1   & 1.2$\pm$0.3 &  $^{13}$CS	- C$^{34}$S  J=2$\rightarrow$1, 3$\rightarrow$2  \\
TMC1-NH3+30   &      11.79   &    15.58   & 10.2$\pm$2.8   &  1.2$\pm$0.8   &     &   &   $^{13}$CS	- C$^{34}$S  J=2$\rightarrow$1, 3$\rightarrow$2      \\   
TMC1-NH3+60   &      12.12   &    12.88   & 12.7$\pm$1.1   &  1.5$\pm$0.8   &     &   &    $^{13}$CS	- C$^{34}$S  J=2$\rightarrow$1, 3$\rightarrow$2          \\ 
TMC1-NH3+120  &      13.10   &    10.04   & 11.9$\pm$1.6   &  0.29$\pm$0.14  &    &   &    C$^{34}$S - CS  J=1$\rightarrow$0, 2$\rightarrow$1, 3$\rightarrow$2  \\ 
TMC1-NH3+180  &      13.78   &     4.04   & 11.4$\pm$1.4   &  0.16$\pm$0.06  &    &   &   CS         J=1$\rightarrow$0, 2$\rightarrow$1, 3$\rightarrow$2 \\  
TMC1-NH3+240  &      13.10   &     2.18   & 13.5$\pm$2.3   &  0.19$\pm$0.19  &    &   &   CS         J=1$\rightarrow$0, 2$\rightarrow$1, 3$\rightarrow$2    \\  \hline
TMC1-C+0     &      11.26   &    19.85   &  8.5$\pm$2.0    &  4.5$\pm$3.4     &   &   &   $^{13}$CS	- C$^{34}$S  J=2$\rightarrow$1, 3$\rightarrow$2 \\
TMC1-C+30    &      11.32   &    18.47   &  10.3$\pm$2.0   &  4.3$\pm$2.3     &   &   &   $^{13}$CS	- C$^{34}$S  J=2$\rightarrow$1, 3$\rightarrow$2 \\
TMC1-C+60    &      11.67   &    13.34   &  11.6$\pm$2.2   &  1.19$\pm$0.45   &   &   &   $^{13}$CS	- C$^{34}$S  J=1$\rightarrow$0, 2$\rightarrow$1, 3$\rightarrow$2 \\
TMC1-C+120   &      13.13   &     4.79   &  11.1$\pm$1.9   &  0.54$\pm$0.25   &   &   &  C$^{34}$S - CS      J=1$\rightarrow$0, 2$\rightarrow$1, 3$\rightarrow$2  \\
TMC1-C+180   &      14.08   &     2.20   &  13.5$\pm$1.1   &  0.55$\pm$0.28   &   &   &   C$^{34}$S - CS      J=1$\rightarrow$0, 2$\rightarrow$1, 3$\rightarrow$2 \\
TMC1-C+240   &      14.53   &     1.63   &  13.5$\pm$2.7   &  0.26$\pm$0.18   &   &   &   CS  J=1$\rightarrow$0, 2$\rightarrow$1, 3$\rightarrow$2  \\

\hline \hline
\end{tabular}
\end{table*}

\section{Physical conditions: Gas kinetic temperature and molecular hydrogen density}

A detailed knowledge of the physical conditions is required for an accurate estimate of the molecular
column densities and abundances. This is specially important in those positions
where we have observed only one line and a multi-transition
study is not possible. In these cases, the knowledge of the gas kinetic 
temperature and density is imperative.
CS is a diatomic molecule with well known collisional coefficients \citep{Denis2018, Lique2006a} 
that has been largely used as density and column density tracer in the interstellar medium. 
Moreover, the velocity-component analysis presented in Sect. 5 shows 
that CS is detected in the 5.5~km~s$^{-1}$ and 6.0 ~km~s$^{-1}$ components, 
which encompass the bulk of the dense molecular gas (compare, e.g., the C$^{18}$O and CS profiles in Fig.~\ref{profiles}).
Therefore, we consider that CS and its isotopologues are good tracers of the 
average physical conditions in this cloud.

In order to derive the gas physical conditions, we fit the line intensities of the 
observed  CS, C$^{34}$S and $^{13}$CS lines using the molecular excitation and
radiative transfer code RADEX \citep{Tak2007}.
During the fitting process,  we fix the isotopic ratios 
to $^{12}$C/$^{13}$C=60, $^{32}$S/$^{34}$S= 22.5 \citep{Gratier2016} and
assume a beam filling factor of 1 for all transitions (the emission is more extended than
the beam size). Then, we let T$_k$, n(H$_2$) and  N(CS) vary as free parameters.
The parameter space (T$_k$, n(H$_2$) and  N(CS)) is then explored following the 
Monte Carlo Markov Chain (MCMC) methodology with a Bayesian inference approach. 
In particular, we used the emcee \citep{Foreman2012}
implementation of the Invariant MCMC Ensemble sampler methods by \citet{Goodman2010}. 
While n(H$_2$) and  N(CS) are allowed to vary freely, we need to use a prior to limit the gas
kinetic temperatures to reasonable values in this cold region and hence  break the temperature-density degeneracy
that is usual in this kind of calculations.

\begin{center}
\begin{figure}
\includegraphics[angle=0,scale=.5]{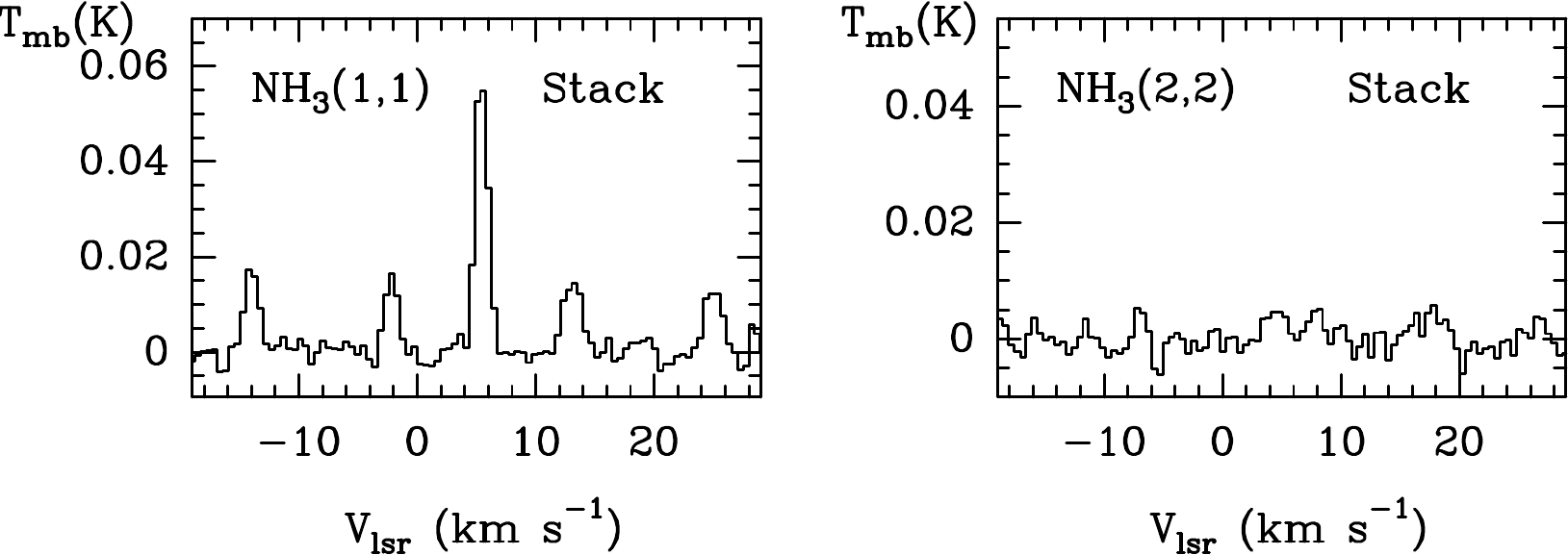}
\caption{NH$_3$ (1,1) and NH$_3$ (2,2) lines profiles obtained by stacking all the band-K
spectra observed with the 40m Yebes telescope (A$_V <$ 10 mag) towards TMC~1.} 
\label{nh3}
\end{figure}
\end{center}

The prior in the gas kinetic temperature is based on our knowledge of the dust temperature
from Herschel maps (Kirk et al., in prep). Gas and dust are expected to be thermalized in regions 
with n(H$_2$)$>$10$^4$ cm$^{-3}$. 
\citet{Friesen2017} estimated the gas kinetic temperature in a wide sample of 
molecular clouds based on the NH$_3$ (1,1) and (2,2) inversion lines,
and obtained that the gas temperature is systematically $\sim$1-2 K lower than the 
dust temperature obtained from the Herschel maps.
This discrepancy is interpreted as the consequence of the single-temperature SED fitting procedure  
which assumes that all the dust 
is at the same temperature along the line of sight. Towards a starless core in which the dust in the
surface is warmer than in the innermost region, this approximation would produce an
overestimation of the dust temperature. In order to account for these effects, in our MCMC calculations 
we use a flat prior to the gas kinetic temperature with constant probability for T$_k$=T$_d$$\pm$5 K
and zero probability outside. 

In Table~3, we show the gas temperature and the density derived from the
multi-line fitting of CS and its isotopologues. Across the cuts,  there are two differentiated 
regions: i) for Av$<$7.5~mag, the density is
quite uniform and similar to a few 10$^3$~cm$^{-3}$ and gas temperatures are about 13-15~K 
which corresponds to gas thermal pressure of $\sim$5 $\times$ 10$^4$~K~cm$^{-3}$. Hereafter, we will refer
to this moderate density envelope as the {\it translucent} component; 
and ii) for Av$>$7.5 mag, the density is an order of magnitude larger, T$_k$$\sim$10 K and the density
keeps increasing towards the extinction peak. 
Hereafter, we will refer to this region as the {\it dense} component. The gas pressure in the dense phase
is about 10 times larger than in the translucent phase. The transition from one phase
to the other occurs in $<$ 60" ($\sim$0.04 pc) and it is not well sampled by our data (see Fig.~\ref{nh2}). 

The low densities found in the translucent phase,  n(H$_2$)$\sim$ a few10$^3$ cm$^{-3}$, might cast 
some doubts about our assumption of  gas and dust thermal equilibrium. In order to check this 
hypothesis, we have tried to independently derive the gas kinetic temperature in this region using
our NH$_3$ data. For that, we have stacked all the NH$_3$ (1,1) and (2,2) spectra obtained with the 
40m Yebes telescope  towards the positions with A$_v <$ 10 mag of the three cuts.
The stacked spectra are shown in Fig.~ \ref{nh3}, with a good detection of the NH$_3$ (1,1) line while the
NH$_3$ (2,2) line remains undetected. Even assuming a density as low as n(H$_2$)=10$^3$ cm$^{-3}$, 
a RADEX calculation shows that the non-detections of the  (2,2) line implies
an upper limit of $<$ 15 K for the gas temperature, slightly lower than the dust temperature. 
It is remarkable that  NH$_3$ is only detected in the 5.5 km s$^{-1}$ velocity component
while CS is detected in the 5.5 km s$^{-1}$ and 6.0 km s$^{-1}$ components. The upper limit
to the gas kinetic temperature is only valid for the  5.5 km s$^{-1}$ component that is 
very likely the densest and coldest component.
We consider, therefore, that the temperature derived from the CS fitting is more adequate for our purposes
and it is used hereafter in the molecular abundance calculations. 
In Sect.~8, we show that the temperatures obtained with our CS fitting are in good agreement with those 
predicted with the Meudon PDR code.

Another important assumption in our density estimate process is that the beam filling factor is 1 for
all the transitions. This is based on the morphology of the Herschel maps that present
smooth and extended emission in the translucent part. This assumption is
questionable towards the extinction peaks where the column density map presents
a steeper gradient (see Fig.~\ref{map}). In Table~3 we also compare our
estimates with previous ones by  \citet{Feher2016}  towards TMC~1-CP and TMC~1-NH3, finding 
excellent agreement.  The good agreement between our estimates and 
those derived from NH$_3$, which are not affected by different beam sizes, suggests
that our assumption is not far from the reality.
The derived densities for the translucent component
are in good agreement with previous estimates of the envelope density by 
\citet{Schnee2010} and  \citet{Lique2006}.

\begin{figure*}
\includegraphics[angle=0,scale=.9]{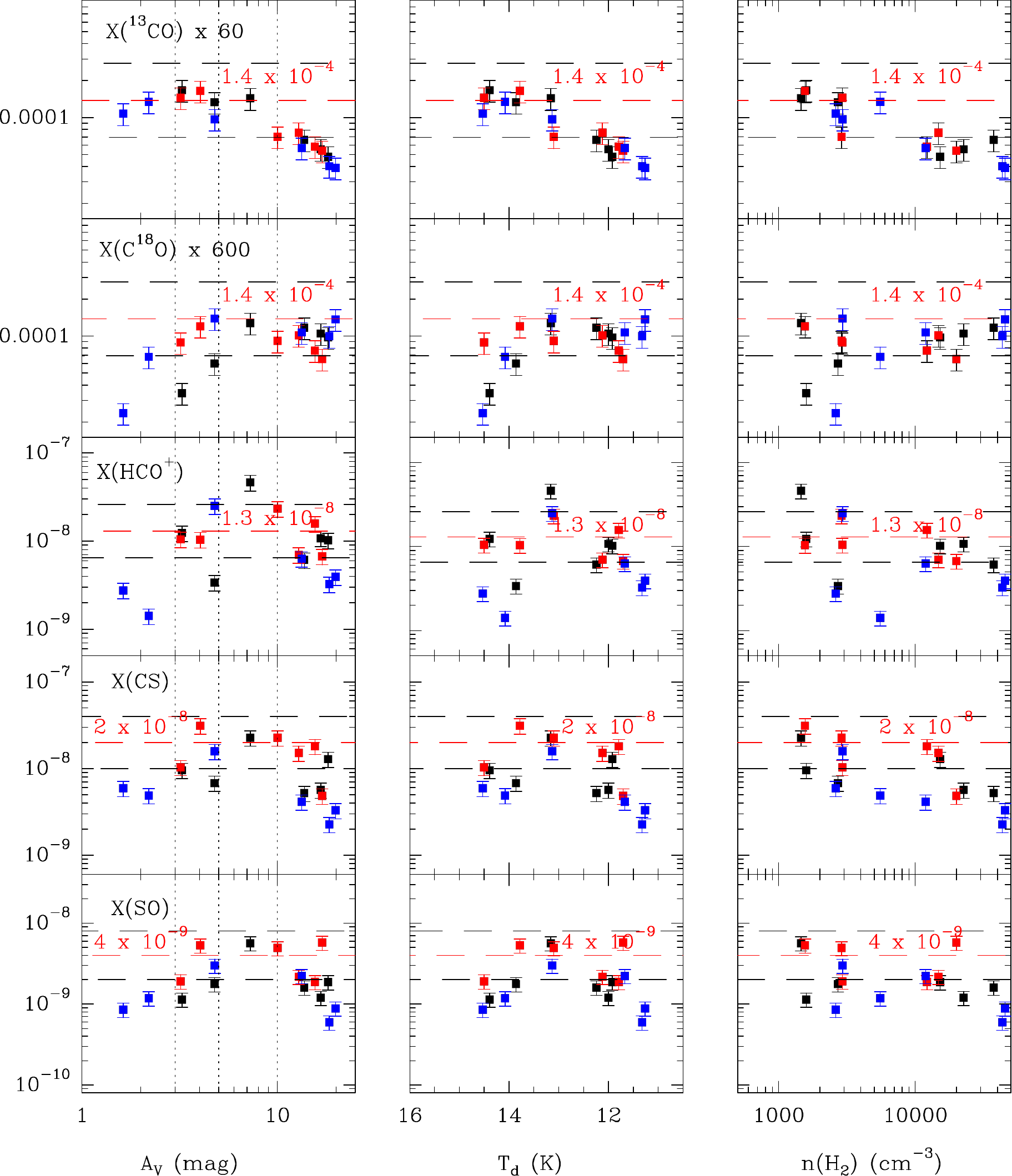}
\caption{Estimated molecular abundances with respect to H$_2$ for the three studied cuts, TMC~1-CP (black squares), TMC~1-NH3 (red) and
TMC~1-C (blue) as a function of the visual extinction (left column), dust temperature (center panel) and molecular
hydrogen densities (right panel). The horizontal lines indicate a representative value
in the translucent part (red dashed line) and a variation of a factor of 2 relative to it (black dashed lines).
Vertical lines mark the A$_V$=3 mag, 5 mag and 10 mag positions.} 
\label{mol1}
\end{figure*}

\begin{figure*}
\includegraphics[angle=0,scale=.9]{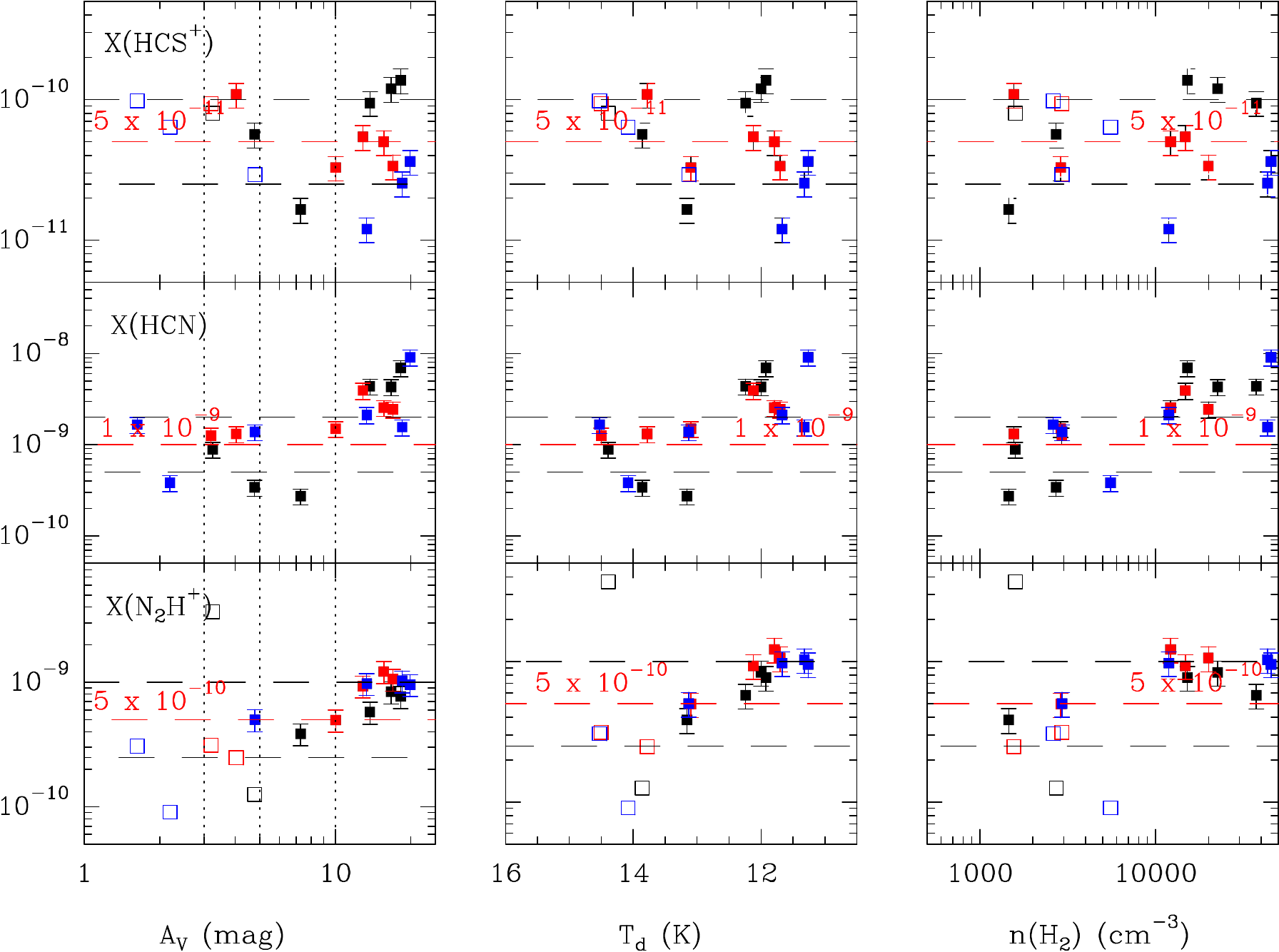}
\caption{Estimated molecular abundances with respect to H$_2$ for the three studied cuts, TMC~1-CP (black squares), TMC~1-NH3 (red) and
TMC~1-C (blue) as a function of the visual extinction (left column), dust temperature (center panel) and molecular
hydrogen densities (right panel). Empty symbols correspond to upper limits. The horizontal lines indicate a representative value
in the translucent part (red dashed line) and a variation of a factor of 2 relative to it (black dashed lines).
Vertical lines mark the A$_V$=3 mag, 5 mag and 10 mag positions.} 
\label{mol2}
\end{figure*}

\begin{table}
\begin{tabular}{lll}\\
\multicolumn{3}{l}{ Table 4: Collisional rate coefficients} \\ \hline \hline
\multicolumn{1}{c}{Mol}     & 
\multicolumn{2}{c}{ref}  \\ \hline \hline                                                  
CO                    &  p-H$_2$   &   \citet{Yang2010}       \\                 
HCO$^+$          &  p-H$_2$      &   \citet{Flower1999}    \\   
HCN                  &  p-H$_2$   &   \citet{Ben2012}        \\                      
CS                     &  p-H$_2$   &   \citet{Denis2018}     \\     
                          &  He            &  \citet{Lique2006a}    \\
SO                     &  p-H$_2$  &   \citet{Lique2007}    \\                  
N$_2$H$^+$     &  p-H$_2$  &   \citet{Daniel2016}     \\
\hline \hline
\end{tabular}
\end{table}

\section{Molecular abundances}
Beam averaged molecular column densities and abundances have been derived using RADEX and
the collisional rate coefficients shown in Table~4. In the following, we add more details of the abundance
calculations.

\subsection{$^{13}$CO, C$^{18}$O}
In this work, we use $^{13}$CO and C$^{18}$O as tracers of CO by assuming fixed
isotopic ratios. We use the physical conditions in Table~3 to derive $^{13}$CO and C$^{18}$O
column densities from the observations of the J=1$\rightarrow$0 rotational line. 
This line is thermalized in the range of densities
considered, therefore we do not expect any uncertainty in the molecular abundances
because of the adopted densities. The major uncertainty comes from the opacity effects and 
possible variations in the CO/C$^{18}$O and   CO/$^{13}$CO ratios.

For A$_V$$>$ 7 mag, the $^{13}$CO 1$\rightarrow$0 line
is expected to be optically thick ($\tau >$ 1). In this region,
we use C$^{18}$O as tracer of the CO abundance by assuming $^{16}$O/$^{18}$O=600 \citep{Wilson1994}.
For A$_V$$<$7 mag, the $^{13}$CO line is expected to be optically thin. In this region we have
separately estimated the $^{13}$CO and C$^{18}$O column densities which allows us to investigate
the N($^{13}$CO)/N(C$^{18}$O) ratio. Interestingly, this ratio increases from $\sim$10 at 
A$_V$$\sim$7 mag to $\sim$40 at A$_V$$\sim$ 3 mag. This result is not the consequence of
the numerous velocity components along the line of sight. As shown in Fig.~\ref{profiles}, the
T$_b$($^{13}$CO 1$\rightarrow$0)/T$_b$(C$^{18}$O 1$\rightarrow$0) is $>$20 in all the velocity
components towards the offset (240",0). Similarly 
T$_b$($^{13}$CO 1$\rightarrow$0)/T$_b$(C$^{18}$O 1$\rightarrow$0)$\sim$4
in all velocity components towards the offset (120",0). The high 
N($^{13}$CO)/N(C$^{18}$O) ratio is more likely the consequence of selective photodissociation 
and isotopic fractionation in the translucent cloud \citep{Liszt1996, Bron2018}. 
Because of isotopic fractionation in the external part of the cloud, a reliable estimate of N($^{12}$CO) 
requires the comparison with a chemical model that  includes a differentiated chemistry for the CO isotopologues. We will discuss
this phenomenon in Sect. 8 when we compare our results with the Meudon PDR code predictions. 

In Fig.~\ref{mol1}, we plot the CO abundance derived as X($^{13}$CO)$\times$60 and  X(C$^{18}$O)$\times$600 
as a function of visual extinction, dust temperature and density in the three observed cuts. The decrease of
the $^{13}$CO  abundance for A$_V$ $>$ 10~mag is not real but the consequence of the high optical depth of the observed lines.
In this dense regions, the rarer isotopologue C$^{18}$O is a better tracer of the CO abundance. 
The C$^{18}$O abundance sharply decreases for A$_V$ $<$ 3~mag. This is consistent 
with the threshold of A$_V$ = 1.5~mag proposed by \citet{Cernicharo1987} for the C$^{18}$O detection. The CO
abundance presents a peak of $\sim$1.4$\times$10$^{-4}$ at 
Av $\sim$ 3 mag (T$_d$=14 K) and seems to decrease towards the dense core phase.
This is the expected behavior in this dark cloud where the dust temperature is lower than the 
evaporation temperature (T$_{evap}$=15-25 K), so that freeze-out is expected. 

\subsection{HCO$^+$, H$^{13}$CO$^+$, HC$^{18}$O$^+$}
We have observed the J=1$\rightarrow$0 rotational lines of HCO$^+$,
H$^{13}$CO$^+$ and HC$^{18}$O$^+$. Column densities of all isotopologues 
have been derived using RADEX and the physical
parameters in Table~3. In our column density calculations, we only 
use the H$^{13}$CO$^+$ and HC$^{18}$O$^+$ spectra 
because the HCO$^+$ J=1$\rightarrow$0 line presents self-absorbed profiles. 
Then, we derive the HCO$^+$ column density from the rarer isotopologues assuming  
N(HCO$^+$)/N(H$^{13}$CO$^+$)=60 or N(HCO$^+$)/N(HC$^{18}$O$^+$)=600. 
The results are shown in Fig.~\ref{mol1}.
The HCO$^+$ abundance is maximum at A$_V$$\sim$5$-$10 mag, i.e., 2 mag deeper than CO 
in the translucent cloud. The abundance of HCO$^+$
further decreases towards the dense high extinctions peaks. 

\subsection{CS, C$^{34}$S, $^{13}$CS}
The CS column densities have been derived as explained in Sect.~6. We find that 
the CS abundance is maximum at A$_V$$\sim$5$-$10 mag, with abundances with respect 
to H$_2$,  X(CS) $\sim$ 2$\times$10$^{-8}$. 
The dispersion in the derived CS abundances in translucent cloud is 
of a factor of 2. For A$_v >$10 mag, the CS abundance sharply decreases suggesting a rapid 
chemical destruction or freeze out on the grain mantles. 

\subsection{SO,$^{34}$SO}
Several lines of SO and $^{34}$SO lie in the frequency range covered by our setups (see Table~B.2).
Regarding $^{34}$SO, only the J=2$_3$$\rightarrow$1$_2$  has been detected towards the high extinction
positions of the TMC~1-CP, TMC~1-NH3 and TMC~1-CP cuts.  Towards the positions where we detect this $^{34}$SO line, 
we measure T$_b$(SO 2$_3$$\rightarrow$1$_2$)/T$_b$($^{34}$SO 2$_3$$\rightarrow$1$_2$)
of 10$-$20. This implies opacities $<$2, i.e., the emission is moderately optically thick in the dense region. 
We have estimated the SO column density based on
the RADEX fitting of the main isotopologue lines in the dense and translucent phases.
Towards the high extinction peaks, we estimate an uncertainty of a factor of 2
in the column density estimates because of the moderate opacity.  The derived SO abundances
are $\sim$1.8 $\times$10$^{-9}$, 0.9 $\times$10$^{-9}$ and 2.9 $\times$10$^{-9}$ for TMC~1-CP, TMC~1-C and TMC~1-NH3, respectively.  
Towards the dense phase,  we have been able to derive the density from the 
 T$_b$(SO 3$_4$$\rightarrow$2$_3$)/T$_b$(SO 2$_3$$\rightarrow$1$_2$) ratio, obtaining values fully consistent
 with those in Table ~3.
Our abundance estimate towards TMC~1-CP is consistent with previous estimates by
 \citet{Ohishi1998}, \citet{Agundez2013} and \citet{Gratier2016}. The overabundance of SO towards TMC~1-NH3 has been already
 pointed out by several authors (\citealp{Lique2006} and references therein).
Similarly to HCO$^+$ and CS, SO presents its maximum abundance at A$_v$$\sim$5-10
with a peak value of $\sim$5$\times$10$^{-9}$. 

\subsection{HCS$^+$}
Because of the weak intensities of the HCS$^+$ lines, we have only detected one line per position which prevents
a multi-transition study. In order to estimate the HCS$^+$ abundances we have assumed the physical conditions in
Table~3 and used the HCO$^+$ collisional rate coefficients. We obtain a large scatter in the HCS$^+$
abundances in the whole range of visual extinctions (see Fig.~\ref{mol2}), without any clear trend of the 
HCS$^+$ abundance with the visual extinction, dust temperature or gas density. Interestingly, we  find 
differences among the HCS$^+$ abundance towards the different cuts, 
being larger towards TMC~1-CP than towards TMC~1-NH3 and TMC~1-C 
(see Fig.~\ref{mol2}).

\subsection{HCN, H$^{13}$CN, HC$^{15}$N, N$_2$H$^+$}
All the N-bearing species included in this subsection share some characteristics:
i) they are only detected in the 5.5~km~s$^{-1}$  component and ii) present larger abundances 
towards the dense phase than towards the outer part of the cloud.

The hyperfine splitting of HCN allows us to estimate the opacity and 
excitation temperature (assuming a beam filling factor of 1 and equal excitation temperature for all
the hyperfine components) and
hence, to obtain an estimate of the density and column density.
In the dense region,  the HCN 1$\rightarrow$0 line presents deep
self-absorption features (see Fig.~A2, A4 and A6). Thus,
we use the isotopologue H$^{13}$CN to calculate the gas density and X(HCN)
assuming N(HCN)/N(HC$^{13}$CN)=60. We
obtain HCN  abundances of 2$-$10 $\times$ 10$^{-9}$ and molecular hydrogen  
densities of a few $\times$ 10$^5$ cm$^{-3}$ in the dense cloud. It is remarkable
that the densities derived from the HCN data are larger than those derived from 
CS and NH$_3$ by a factor of $>$10. 

Relatively intense emission of the  HCN 1$\rightarrow$0 line is detected in the translucent cloud.
Because of the large dipole moment of HCN, molecular hydrogen densities $>$10$^4$ cm$^{-3}$ 
are required to achieve excitation temperatures $>$ 5 K and hence detectable emission.
One possibility is that the densities in the 5.5~km~s$^{-1}$ component of the translucent cloud
are n(H$_2$)$>$10$^4$ cm$^{-3}$. In fact, following the procedure described
above, we derive densities $\sim$a few 10$^4$ cm$^{-3}$ and 
X(HCN)$\sim$1$\times$10$^{-9}$  (see Fig.~\ref{mol2}) in the 3$<$A$_V$$<$10 mag range.
Alternatively, radiative trapping could have an important role 
in the excitation of the  HCN 1$\rightarrow$0 line in TMC~1. Radiative excitation would explain the 
excitation of the HCN line without invoking higher densities in the 5.5~km~s$^{-1}$ velocity component .
Assuming a core/envelope system (two-phase model) and using a Monte Carlo radiative transfer code, \citet{Gonzalez1993}  explained
the line intensities in TMC~1 with HCN abundances of $\sim$5$ \times$10$^{-9}$ all across the cloud. 
In this two-phase model, the HCN molecules in the envelope are excited by the photons coming 
from the core which is a bright source in the HCN 1$\rightarrow$0 line without the need of 
invoking higher densities. The value thus derived by \citet{Gonzalez1993} is similar to that we have
derived for the dense gas and a factor of $\sim$5 larger than the HCN abundances  we
derive in the translucent cloud. Comparing with these results, we consider that our estimates of 
the HCN abundance based on one-phase molecular excitation calculations 
are accurate within a factor of 5$-$10.

Similarly to HCN,  we have fitted the 
excitation temperature and opacity of the N$_2$H$^+$ 1$\rightarrow$0 line based on the 
hyperfine splitting, and hence the total N$_2$H$^+$ column densities. This molecular ion
has been almost exclusively detected towards the dense region with  abundances of  about 1$\times$10$^{-9}$ 
and densities of a few 10$^4$ cm$^{-3}$, quite consistent with those derived from CS. Towards the translucent cloud, 
we have detections only for A$_V$$>5$ mag with abundances of $\sim$5$\times$10$^{-10}$. 

\section{Chemical modeling of the translucent cloud}

The increase in dust temperature at the edges of molecular clouds is 
understood as the consequence of dust heating by
the surrounding interstellar radiation field (IRSF). Moreover, the high N($^{13}$CO)/N(C$^{18}$O) 
ratio measured in TMC~1 testifies that UV radiation has an active role in the molecular
chemistry. To determine the value of the ambient UV field is hence a requisite for
the appropriate chemical modeling of the region.

\begin{figure}
\includegraphics[angle=0,scale=.6]{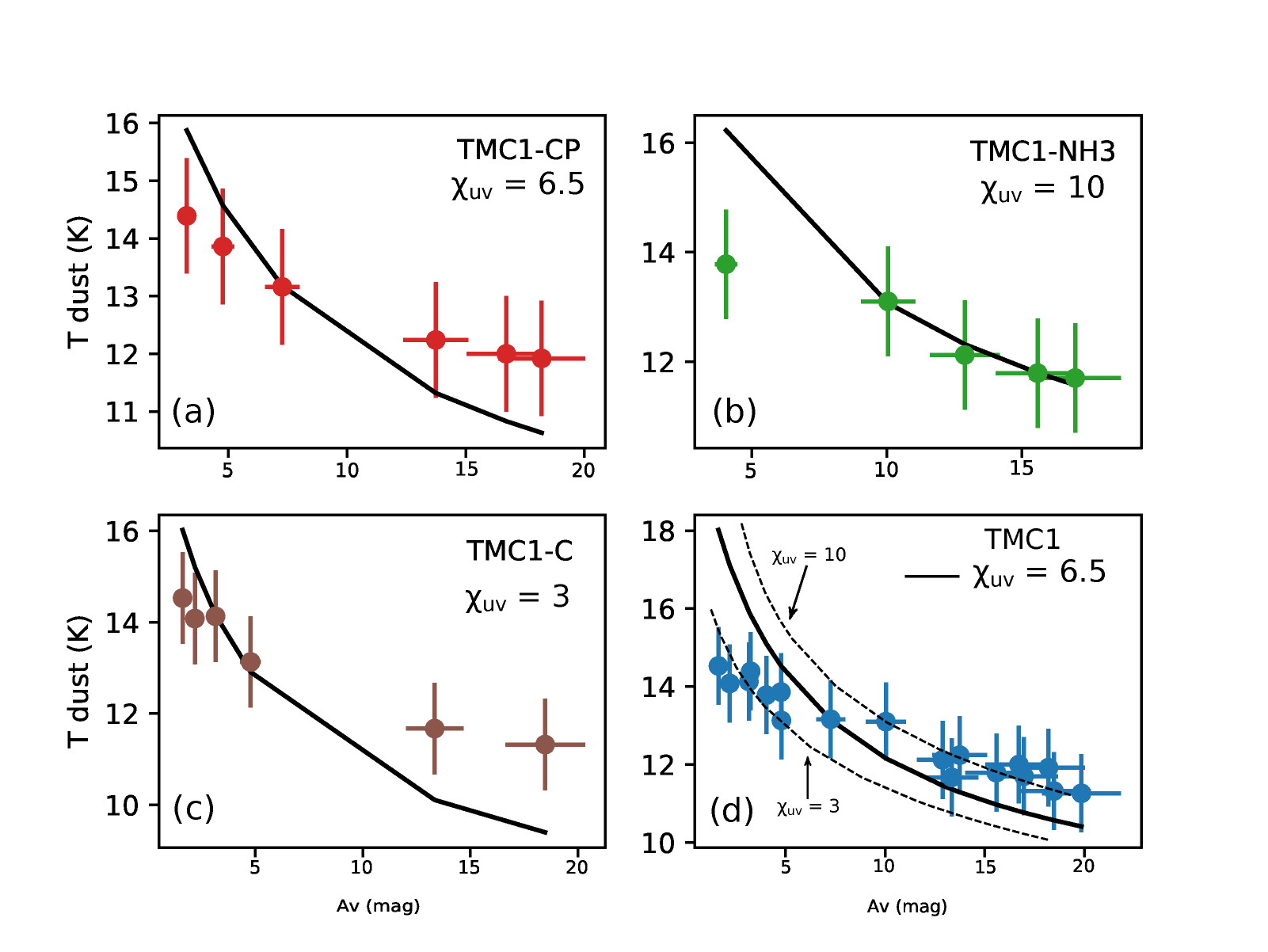}
\caption{T$_d$-A$_V$ fit following the parametric expression published by \citet{Hocuk2017} for the 
observed cuts: TMC~1-CP in {(a)}, TMC~1-NH3 in {(b)}, and TMC~1-C in {(c)}.
In panel {(d)}, we try to fit all the positions from the three cuts together. For A$_{V}<7.5$~mag, the best fit is 
found for $\chi_{UV}\sim 3$. For A$_{V}>7.5$~mag, the best fit corresponds to $\chi_{UV}\sim 10$. 
The average value of $\chi_{UV}\sim 6.5$ is found to provide the best fit for the whole TMC~1 region. } 
\label{draine}
\end{figure}

\begin{table}
\begin{tabular}{lllll}\\
\multicolumn{5}{l}{Table 5 .- PDR chemical models} \\ \hline \hline
\multicolumn{1}{c}{}     & 
\multicolumn{1}{c}{$\zeta_{\rm H_2}$ (s$^{-1}$)} & \multicolumn{1}{c}{C/H} & \multicolumn{1}{c}{C/O} 
&  \multicolumn{1}{c}{S/H} \\ \hline \hline                                                  
A            &    5$\times 10^{-17}$   &  1.38$\times 10^{-4}$  &   0.4    &  1.5$\times 10^{-5}$   \\     
B            &    5$\times 10^{-17}$   &  7.90$\times 10^{-5}$  &   0.4    &  1.5$\times 10^{-5}$    \\
C            &    5$\times 10^{-17}$   &  3.90$\times 10^{-5}$  &   0.4    &  1.5$\times 10^{-5}$    \\
D            &    5$\times 10^{-18}$   &  1.38$\times 10^{-4}$  &   0.4    &  1.5$\times 10^{-5}$    \\
E            &    1$\times 10^{-16}$   &  1.38$\times 10^{-4}$  &   0.4    &  1.5$\times 10^{-5}$    \\
F            &    5$\times 10^{-17}$   &  1.38$\times 10^{-4}$  &   1.0    &  1.5$\times 10^{-5}$   \\
G            &    5$\times 10^{-17}$   &  1.38$\times 10^{-4}$  &   0.8    &  1.5$\times 10^{-5}$    \\
H            &    5$\times 10^{-17}$   &  7.90$\times 10^{-5}$  &   1.0    &  1.5$\times 10^{-5}$    \\
I            &    5$\times 10^{-17}$   &  7.90$\times 10^{-5}$  &   1.0    &  8.0$\times 10^{-7}$    \\
Best-fit         &    5$\times 10^{-17}$   &  7.90$\times 10^{-5}$  &   1.0    &  8.0$\times 10^{-7}$    \\
\hline \hline
\end{tabular}
\end{table}

\begin{figure*}
\includegraphics[angle=0,scale=.9]{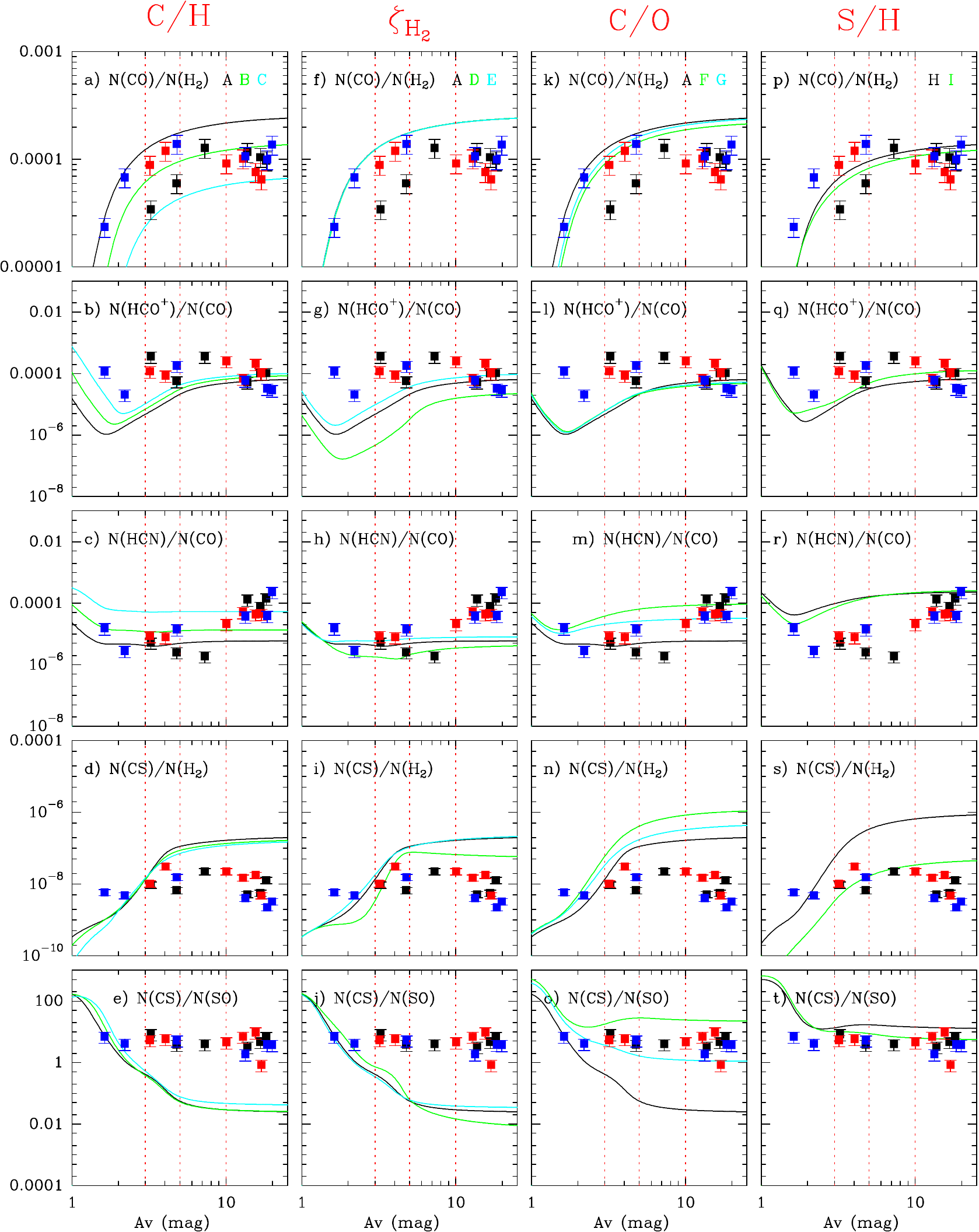}
\caption{Comparison between the predictions of models listed in Table~5 and the
cumulative column densities derived in TMC~1. In this Figure, we have selected N(CO)/N(H$_2$), N(HCO$^+$)/N(CO),
N(HCN)/N(CO), N(CS)/N(H$_2$) and N(CS)/N(SO) to explore the parameter space, where N(CO) has been derived from
our observations as N(CO)=600$\times$N(C$^{18}$O) and N(H$_2$)=A$_V$(mag)$\times$10$^{21}$ cm$^{-2}$.
The observational points are indicated with squares and different colors correspond to the three observed
cuts as follows: black for TMC~1-CP, red for TMC~1-NH3 and blue for TMC~1-C. Dashed red lines indicate
A$_V$ = 3 mag (C$^+$/C/CO transition region), A$_V$ = 5 (translucent cloud) and A$_V$ =10 mag (dense region).} 
\label{chem1}
\end{figure*}

\subsection{Estimate of the incident UV field in TMC~1: Dust temperature}
Herschel has provided extensive H$_2$ column density and dust temperature maps,
with high angular resolution ($\sim$36$"$), of nearby star-forming regions. These maps
constitute an unprecedented opportunity to determine the incident UV field and its
local variations due to the nearby star formation activity.
Dust temperatures are established by the radiative equilibrium balance between the 
absorption of UV/visible photons and the emission at a given temperature, T$_d$. In the cloud
border, the exact value of T$_d$ depends on the local IRSF and the absorption efficiencies of grains that
are dependent on the grain composition and size. Deeper in the cloud, grain heating is produced by 
near and mid-infrared emission  coming from the warm dust layer at the surface of the cloud \citep{Zucconi2001}.
The direct calculation of the local UV field as
a function of the dust temperature is hampered by our poor knowledge of the grain composition 
and its detailed variation across the cloud. Another fundamental problem is that in our part of the Galaxy
the dust heating is dominated by the visible part of the ISRF, and  
the visible/IR part of the ISRF does not scale in a simple way with the ultraviolet part. 
Hence, the UV field derived from the dust temperature can only be considered as a 
a first guess of the local UV flux.

Alternatively, we can try to fit the observational data 
with a simple analytical expression. Different attempts have been done to derive 
parametric expressions that relates the UV ambient field and the dust temperature as a function of the visual extinction
\citep{Hollenbach1991,Zucconi2001,Garrod2011,Hocuk2017}.
Most of them provide a good fitting of the observed T$_d$ as a function of the incident UV field, $\chi_{UV}$,
in a given range of visual extinctions but have problems to fit the 
whole range, from A$_V$=0.01 to A$_V$$>$50 mag.
We have used the most recent parametric expression by \citet{Hocuk2017}
to obtain an estimate of the incident UV field.
This expression is well adapted to the range of visual extinctions relevant to 
this paper (3 mag $<$ A$_V$ $<$ 20 mag) and is consistent with what one would
expect for a mixed carbonaceous-silicate bared grains. 
\begin{equation}
T_d= [11 + 5.7 \times {\rm tanh}(0.61 - {\rm log10} (A_V))] \chi_{UV}^{1/5.9}
\end{equation}
where $\chi_{UV}$ is the UV field in Draine units and the the visible/IR part of
the IRSF is assumed to scale with  $\chi_{UV}$.

Fig.~\ref{draine} shows the T$_d$-A$_V$ plots for the 3 cuts considered in this paper.
None of the cuts can be fitted with a single value of the UV field.
In fact, the dense cloud (A$_V$$>$ 7.5 mag) is better fitted with  $\chi_{UV}$$\sim$10 
while the translucent cloud is fitted with $\chi_{UV}$$\sim$ 3. Moreover, the three observed cuts
share the same behavior without any hint of variation of $\chi_{UV}$ from one cut to another
(see Fig.~\ref{draine}). 
One compelling possibility is that this break at A$_V$$\sim$ 7.5 mag is caused by a change
in the grain properties. A thick layer of ice would allow the dust to be warmer by up to 15\% 
at  visual extinctions $>$ 10 mag, i.e., the dense component \citep{Hocuk2017}. 
In fact, if we decrease the dust temperature in the dense regions by this factor, we could explain all the positions with
$\chi_{UV}$$\sim$3. This interpretation is also consistent with the sharply decrease in the abundances of the
C- and S- bearing molecules with visual extinction in the dense phase. 
We cannot discard, however, local variations  with the local UV field being lower in 
the northern cut (TMC~1-C) than in the southern cuts TMC~1-CP and  TMC~1-CP.
As commented above, this is a first guess of the local $\chi_{UV}$. In next section,
we will confirm the derived values of $\chi_{UV}$ by comparing gas temperatures and chemical abundances with 
the predictions of the Meudon PDR code.

\begin{figure}
\includegraphics[angle=0,scale=.9]{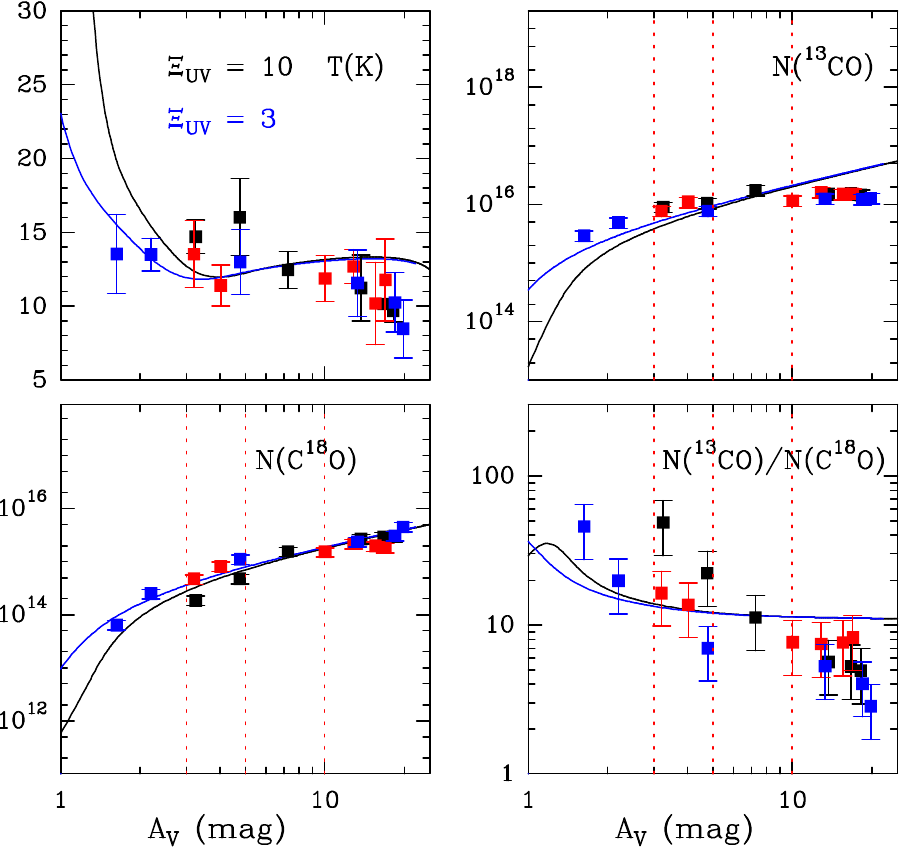}
\caption{Comparison between our "Best-fit" model (see Table~5) and the
molecular abundances derived in this work. The black line corresponds to 
the side with $\chi_{UV}$=10 and the blue line to  $\chi_{UV}$=3.
Dashed red lines indicate A$_V$=3 mag, 5 mag and 10 mag. The observational
points are indicated with squares as in Fig.~\ref{chem1}. }
\label{best-co}
\end{figure}

\begin{figure*}
\includegraphics[angle=0,scale=.95]{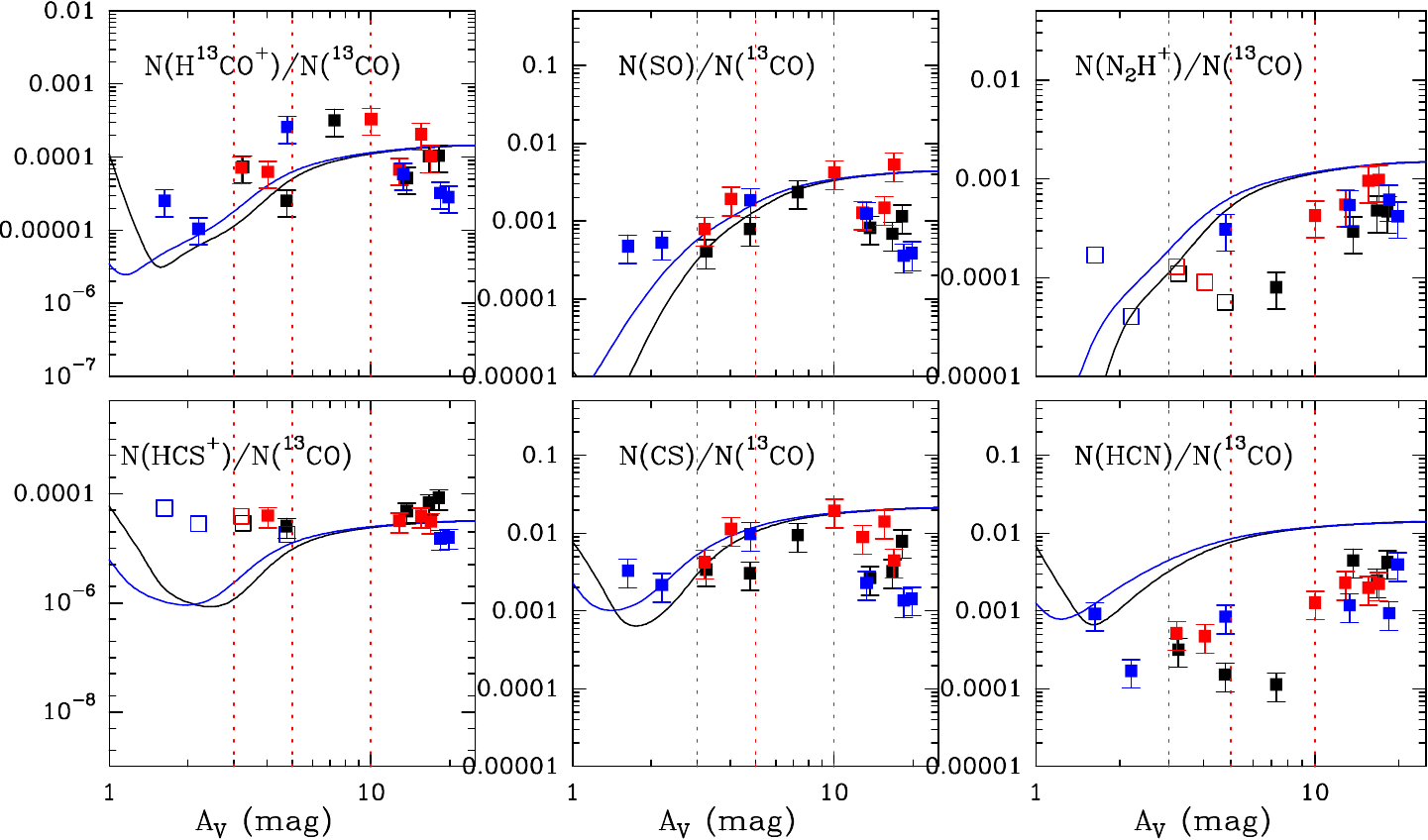}
\caption{Comparison between our "Best-fit" model (see Table~5) and the
molecular abundances derived in this work. The black line corresponds to 
the side with $\chi_{UV}$=10 and the blue line to  $\chi_{UV}$=3.
Dashed red lines indicate A$_V$=3 mag, 5 mag and 10 mag. The observational
points are indicated with squares as in Fig.~\ref{chem1}. }
\label{best-iso}
\end{figure*}

\subsection{Chemical code}

We use the steady-state gas-phase Meudon PDR code 1.5.2 
\citep{Lepetit2006, Goicoechea2007, Gonzalez2008, Bourlot2012}  to estimate the C, O and S
elemental abundances. This code computes the steady-state solution to the thermal balance
and gas-phase chemical network using accurate radiative transfer calculations and a 
plane-parallel geometry. The model explicitly considers the H$_2$ formation on the grain surfaces and adsorption/desorption 
of H and H$_2$ from grains. It does not include the accretion/desorption mechanisms for other  gas-phase molecules.
We assume gas phase elemental abundances below the solar values to take approximately into account depletion effects.
The Meudon PDR code uses an extensive gas-phase chemical 
network which includes the reactions for the $^{13}$C and $^{18}$O isotologues
allowing the direct comparison of chemical predictions with the 
observed $^{13}$CO and C$^{18}$O column densities.

As a first step, we run a series of models in order to explore the parameter space to
determine the values of the cosmic ray ionization rate and elemental abundances that best
fit our observations  (see Table~5). The adopted physical structure is based 
on our previous calculations, the 'a priori' knowledge of
the source (see \citealp{Ebisawa2019} and references therein) and the assumption
of pressure equilibrium. In practice, we carried out the calculations for 
an isobaric plane-parallel 30~mag cloud with a constant 
pressure of 5$\times$10$^4$ K cm$^{-3}$, consistent with the physical parameters
derived from our observations.  Our cloud is illuminated by a UV field, 
$\chi_{UV}^{front}$,  from the front side and $\chi_{UV} =$1 from the back. We consider two
values of  $\chi_{UV}^{front}$,   $\chi_{UV}^{front}$=10 and 3, which is the range of values derived in Section 8.1.
This simple plane-parallel geometry mimics the scenario of a compressed gas layer illuminated from the front
proposed by \citet{Ebisawa2019} on the basis of OH 18cm observations.

Fig.~\ref{chem1} shows the predicted 
N(CO)/N(H$_2$), N(HCO$^+$)/N(CO), N(CS)/N(H$_2$), and N(CS)/N(SO) ratio as a function of the visual extinction 
for  $\chi_{UV} =$10, where N(X) is the cumulative column density from 0 to A$_V$ mag of 
the species X. In general, any line of sight passes through the illuminated cloud surface and the cumulative
column density is the parameter directly related with the observed line intensities.
In the first column, we  show the behavior of the observed abundances 
under changes in C/H. Here we compare the predicted
cumulative column density N(CO) with the observed column density of C$^{18}$O$\times$600
to avoid isotopic fractionation effects. 
For A$_{V} >$3 mag, all the carbon atoms are basically locked
in CO and C/H is well determined from the measured CO abundance. 
Our peak CO abundance, $\sim$1.4 $\times$ 10$^{-4}$, shows that even at the 
low extinction of A$_V$ = 3 mag, carbon is depleted by a factor 
of $\sim$2. Beyond  A$_V$ = 3 mag, the carbon depletion progressively increases
to reach values of $\sim$3 at A$_V$$\sim$10 mag.

In the second column, we investigate the chemical effect of varying  $\zeta(H_2)$. 
The value of $\zeta(H_2)$ mainly affects the predicted N(CS) and the N(HCO$^+$)/N(CO) abundance
ratios. The abundance of CS is also dependent on the value of S/H (third column of Fig.~\ref{chem1}) 
which justifies to use  N(HCO$^+$)/N(CO) as a probe of  $\zeta(H_2)$. 
Values of $\zeta(H_2)$ $>$(5$-$10) $\times$10$^{-17}$ s$^{-1}$ are required to fit the observed
values of the N(HCO$^+$)/N(CO) ratio in TMC~1.  The fit is, however, not perfect with several points
being over model predictions for all the considered values of $\zeta(H_2)$. We will
discuss the value of $\zeta(H_2)$ in more details in Sect 8.5.  

The N(HCN)/N(CO) as well as N(CS)/N(SO) abundance ratios are highly dependent 
on the C/O gas phase elemental ratio (third column of Fig.~\ref{chem1}). We have not 
been able to fit both ratios with a single C/O value. 
While the observed N(HCN)/N(CO) is well explained with C/O$\sim$0.4, N(CS)/N(SO) points to a 
value of the C/O $\sim$0.8$-$1. As commented in Sect. 4.5,  the HCN abundance might be underestimated in this
moderate density gas. Therefore, we consider that N(CS)/N(SO) is a more reliable tracer of the C/O ratio.
We note also that the N(CS)/N(SO) ratio is not very dependent on the S/H value (column 4 of
Fig.~\ref{chem1})  which supports the usage of this ratio to determine C/O. 
Interestingly, there is not a systematic trend of the N(CS)/N(SO) ratio in the translucent cloud that could be identified with 
a preferential oxygen depletion in this range of visual extinctions. Our data are best fit with C/O$\sim$0.8-1.
Once the values of $\zeta(H_2)$ and C/O are fixed, the abundances of CS and SO depend almost linearly 
with S/H (column 4). The abundance of CS is well fitted with  S/H$\sim$8$\times$10$^{-7}$, i.e., a factor of 20
lower than the solar value, S/H $\sim$1.5$\times$10$^{-5}$. 

\begin{figure*}
\includegraphics[angle=0,scale=.9]{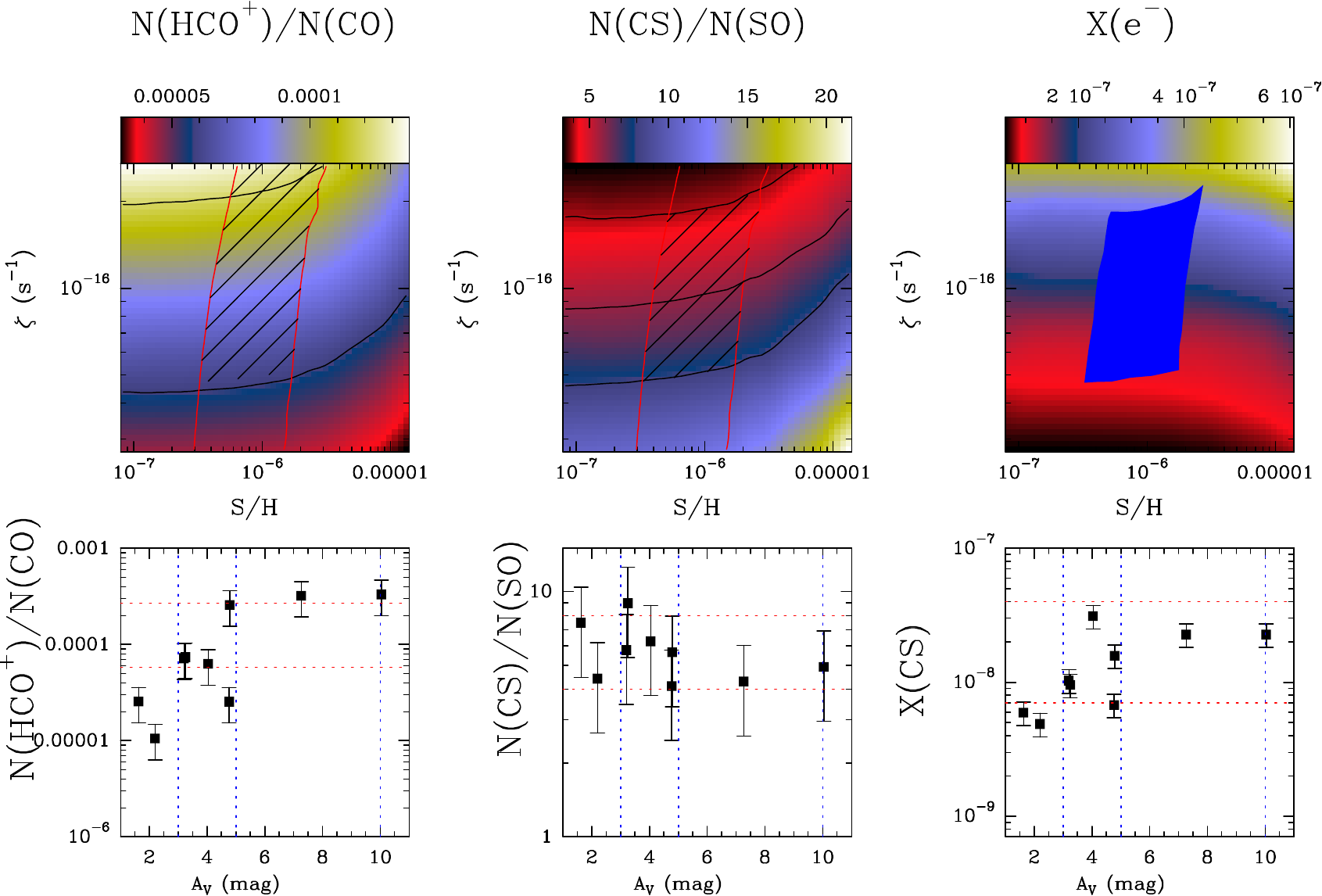}
\caption{{\em Left:} The color map shows the N(HCO$^+$)/N(CO) ratio as a function of $\zeta_{\rm H_2}$ and S/H. Black contours 
correspond to N(HCO$^+$)/N(CO)=5.8$\times$10$^{-5}$ and 1.2$\times$10$^{-4}$. 
Red contours indicate X(CS)=1$\times$10$^{-8}$ and 4$\times$10$^{-8}$, which is the range of CS abundances derived
in the translucent part. {\em Center:} The color map shows the N(CS)/N(SO) ratio as a function of $\zeta_{\rm H_2}$ and S/H. Contours 
are  N(CS)/N(SO)=4, 5.8, and 8. Red contours are the same as in the left panel. {\em Right:} Gas ionization 
fraction, X(e$^-$)=n(e$^-$)/n$_{\rm H}$, as a function of  $\zeta_{\rm H_2}$ and S/H. According with our data, the blue region marks 
the range of values expected in the translucent part of TMC~1. Bottom panels show the observed values in the translucent cloud. Dashed red lines are:
 5.8$\times$10$^{-5}$ and 2.7$\times$10$^{-4}$ in the N(HCO$^+$)/N(CO) panel; 4 and 8 in the N(CS)/N(SO) panel and
 7$\times$10$^{-9}$ and 4$\times$10$^{-8}$ in the X(CS) panel.}
\label{xe}
\end{figure*}

Following the analysis described above, we propose that  C/H$\sim$7.9$\times$10$^{-5}$, C/O$\sim$1,
and S/H$\sim$8$\times$10$^{-7}$, are the most likely values in the translucent cloud, hereafter our
"Best-fit" model.  In the following we extensively compare this model with our observations.

\subsection{C depletion and gas temperature}
The gas kinetic temperature is determined by the balance between the heating and cooling processes at
a given distance from the cloud border.  In molecular clouds, CO is the main coolant
and carbon depletion is determining the gas temperature. The CO lines are optically thick even  at moderate
visual extinctions and the CO line profiles usually present self-absorption features. For this reason,
we have used the rarer isotopologue C$^{18}$O to estimate the total CO
column density by assuming a fixed N($^{12}$CO)/N(C$^{18}$O) ratio.
The CO abundances thus obtained were used to compare with observations in Fig.~\ref{chem1}.

The Meudon PDR code computes the abundances of CO, $^{13}$CO and C$^{18}$O taking into account 
the self-shielding effects for $^{12}$CO  and $^{13}$CO, and the effects of the possible overlap 
between H, H$_2$, CO and $^{13}$CO UV transitions, as well as the isotopic fractionation reactions. Here, we 
directly compare the predicted $^{13}$CO and C$^{18}$O abundances  with observations.
In Fig.~\ref{best-co} we show  our "best-fit" model with $\chi_{UV}$=10 (black) and $\chi_{UV}$=3 (blue) 
predictions with the observed $^{13}$CO and C$^{18}$O column densities. 
For further comparison, we also plot the gas temperature as computed
by the Meudon code. We find quite good agreement of the observed T$_{gas}$ and N(C$^{18}$O)
in the TMC~1-CP and TMC~1-NH3 cuts for $\chi_{UV}$=10 and 3. In these cuts, we do not have
any observed position with A$_V$$<$ 3 mag. On the other hand, the gas kinetic temperature 
in the cut across TMC~1-C are better fitted with $\chi_{UV}$=3.
These results support the interpretation of a lower UV field towards in the northern part of the filament 
than towards the south. In this plot we also represent N($^{13}$CO) and the 
N($^{13}$CO)/N(C$^{18}$O) ratio as a function of the visual extinction from the cloud border.
Although we have a good qualitative agreement between the derived N($^{13}$CO) and observations,
the model underestimates N($^{13}$CO) in the outer A$_V$ $<$ 5~mag. We consider that this disagreement is due
to the fact that, at these low visual extinctions, N($^{13}$CO)
is very sensitive to the uncertainties in the UV illuminating field and the detailed cloud geometry. 

\subsection{Depletions of O and S}
The determination of the O and S depletions is very likely the most challenging
part of this project. The main oxygen reservoirs, O, H$_2$O and OH, cannot be observed in the millimeter domain 
and oxygen depletion has to be derived indirectly from the C/O ratio. As commented above N(HCN)/N(CO) and
N(CS)/N(SO) are good tracers of the C/O ratio. Since we have problems to account for observed abundances 
of HCN and N$_2$H$^+$, we prefer to use N(CS)/N(SO) as a proxy of the C/O ratio.

Atomic S in the outer layers of the cloud
and solid-H$_2$S  in the dense cores are predicted to be the main sulfur reservoirs and both of them are difficult to
observe. Thus, we need to derive the S abundance indirectly by observing minor sulfur compounds. 
The most abundant easily observable S-species in gas phase are CS, SO and H$_2$S. 
Out of them, only CS and SO are thought to be formed in gas phase while the formation of H$_2$S is only understood as 
the product of the hydrogenation of S on the grain surfaces.
Our team has been working in improving the sulfur chemical network by revising the rates of important reactions in 
the CS and SO chemistry at low temperatures. \citet{Fuente2016} already presented calculations of 
the S + O$_2$ $\rightarrow$ SO + O rate at temperatures $<$50~K.  
In this paper, we include new calculations of the SO + OH $\rightarrow$ SO$_2$ + O reaction rate for 
temperatures below 300~K (see Appendix~A). These new rates have been included in the gas-phase chemical network used by the
Meudon code to obtain the most accurate determination of the amount of sulfur in gas phase. 
Fig.~\ref{best-iso} shows the comparison of the CS, SO, HCS$^+$, HCN, and N$_2$H$^+$ column densities with 
model predictions. We have a fair  agreement for all the species except for the N-bearing compounds HCN and N$_2$H$^+$
usingh C/O=1 and S/H=8$\times$10$^{-7}$.We also include HCS$^+$ in our comparison although
we have very few detections of this species in the translucent phase.

\subsection{Gas ionization fraction, X(e$^{-}$)}

The gas ionization fraction determines the coupling of the gas dynamics with the magnetic field
and it is, therefore, a key parameter in star formation studies.  At the scale of the molecular cloud, UV photons 
and cosmic rays are the  main ionization agents in the diffuse/translucent phase, and carbon and sulfur are the main 
electron donors.  At the scale of protostellar disk formation, the dust is the main contributor 
to the magneto-hydrodynamics resistivity. In the following, we discuss the uncertainties in our estimate
of X(e$^-$) and the implications of our results in this context. 

Over the past three decades, several attempts have been carried out to estimate $\zeta_{\rm H_2}$ 
in dense cores from measurements of the abundances of various chemical 
species  (see compilation by  \citealp{Padovani2009}). The values of $\zeta_{\rm H_2}$
derived by \citet{Caselli1998} through DCO$^+$ and HCO$^+$ 
abundance ratios span a range of about two orders of magnitudes 
from $\sim$10$^{-17}$ s$^{-1}$ to $\sim$ 10$^{-15}$ s$^{-1}$.
This large scatter  may  reflect intrinsic variations of the cosmic rays flux from core to core but, as discussed
by the authors, might be the consequence of the sensitivity of the results to several model assumptions, 
mainly the value of specific chemical  reaction rates and elemental depletions. 
In order to minimize these uncertainties, \citet{Fuente2016} used  a gas-phase chemical model to fit the abundances
of  22 neutral and ionic species in order to determine the local value of the cosmic ray ionization rate together with 
the depletion factors towards Barnard 1b. They estimated  a value of $\zeta_{\rm H_2}$ between 
3$\times$10$^{-17}$ s$^{-1}$ and 10$^{-16}$ s$^{-1}$, i.e., an uncertainty of a factor of 4.

Several works have been dedicated to estimate  $\zeta_{\rm H_2}$  in diffuse clouds.
The discovery of significant abundances of H$_3^+$ in diffuse clouds by \citet{Call1998} led 
to values of $\zeta_{\rm H_2}$ $>$ 2 $\times$ 10$^{-16}$ s$^{-1}$. Given
the simplicity of H$_3^+$ chemistry, this value was widely accepted as considered more 
reliable than previous estimates. \citet{Neufeld2010} 
found $\zeta_{\rm H_2}$ $\approx$ 1.2 to 4.8$\times$10$^{-16}$ s$^{-1}$ 
from observations of OH$^+$ and H$_2$O$^+$in clouds with low molecular fraction.
A recent comprehensive work by \citet{Neufeld2017} established that
the ionization rate per H$_2$ in diffuse molecular gas is $\zeta_{\rm H_2}$ = (5.3$\pm$1.1)$\times$10$^{-16}$ s$^{-1}$.

In this work we have estimated the elemental abundances and X(e$^-$) in the translucent part of TMC~1. Our observations
highlight the low-density ($\sim$ a few 10$^3$ cm$^{-3}$) cloud envelope, i.e., the transition 
from the diffuse medium to the dense cores. In the translucent envelope where most of the carbon is already in molecular form,
X(e$^-$) is mainly dependent on $\zeta_{\rm H_2}$ and the elemental abundance of S. 
In order to obtain the range of values of $\zeta_{\rm H_2}$ and S/H consistent with our observations we have run a grid of
models with $\zeta_{\rm H_2}$ = (1, 5.1, 8.6, 14.4, 24.4) $\times$10$^{-16}$ s$^{-1}$ and
S/H=(0.01, 0.018, 0.058, 0.10, 0.19, 0.34, 0.61 and 1.1) $\times$10$^{-5}$. All the other parameters are fixed to the values in
the "Best-fit" model and $\chi_{UV}$=3. We select the model predictions and observations at A$_V$=5 mag, 
as representative of the translucent phase. From our observations, 
we have derived values of N(HCO$^+$)/N(CO) =0.2 $-$ 2.7 $\times$ 10$^{-4}$ at this visual extinction (see Fig.~\ref{xe}). 
Values of  N(HCO$^+$)/N(CO) $>$ 1.3 $\times$ 10$^{-4}$ cannot be fitted with our grid of models (see Fig.~\ref{xe}). 
It is noticeable that an increase of $\zeta_{\rm H_2}$ of a factor of 10,  from $\sim$3 $\times$10$^{-17}$ s$^{-1}$  to  
$\sim$2.2 $\times$10$^{-16}$ s$^{-1}$, only produces an increase of a factor 
of $\sim$2$-$3, from $\sim$5 $\times$ 10$^{-5}$ to  $\sim$1.2 $\times$ 10$^{-4}$, in the
predicted N(HCO$^+$)/N(CO) ratio which makes it very difficult to estimate the value
of $\zeta_{\rm H_2}$ with an accuracy better than a factor of 10 based only on the N(HCO$^+$)/N(CO) ratio.
To further constrain the values of $\zeta_{\rm H_2}$ and S/H, we have tried to fit X(CS) and N(CS)/N(SO) ratio, as well.  
The range of the observed values of N(CS)/N(SO) is of $\sim$4 to 8 which is best fitted 
with $\zeta_{\rm H_2}$ of $\sim$ 0.5 $-$ 1.8 $\times$10$^{-16}$ s$^{-1}$ and S/H $\sim$ 0.4$-$2.2 $\times$10$^{-6}$ 
(see Fig.~\ref{xe}).  Taking all into account, we conclude that $\zeta_{\rm H_2}$ = 0.5 - 1.8 $\times$10$^{-16}$ s$^{-1}$ in the TMC~1
translucent envelope. This value is similar to that found by \citet{Fuente2016} in Barnard 1b and it is within the 
range of the values obtained by \citet{Caselli1998} in dark cores. It is also consistent with the accepted value
in diffuse clouds \citep{Call1998,Neufeld2010}. Therefore, we do not add any evidence of variation 
of $\zeta_{\rm H_2}$ within a factor of 3.

It is also interesting to find the relationship between $\zeta_{\rm H_2}$ and X(e$^-$), that is 
dependent on the local density and the elemental abundances. With the physical and chemical conditions derived in the
studied region,  we need S/H =0.04 $-$ 0.2 $\times$10$^{-5}$ to account for the 
measured CS abundance (see Fig.~\ref{xe}a). As shown in Fig.~\ref{xe}, this would imply that 
X(e$^-$) $\sim$ 9.8 $\times$ 10$^{-8}$ - 3.6 $\times$ 10$^{-7}$ in the transition 
from the diffuse to the dense medium in TMC~1.

\begin{table}
\begin{tabular}{llll}\\
\multicolumn{4}{l}{ Table 6: Molecular abundances and depletions in TMC~1} \\ \hline \hline
\multicolumn{1}{c}{Mol}     & 
\multicolumn{1}{c}{TMC~1-CP} & \multicolumn{1}{c}{TMC~1-NH3} & \multicolumn{1}{c}{TMC~1-C}  \\ \hline \hline                                                  
X(CO)                    &     9.7$\times 10^{-5}$    &    6.5$\times 10^{-5}$    &    1.4$\times 10^{-4}$   \\                 
X(HCO$^+$)          &     1.0$\times 10^{-8}$    &    6.7$\times 10^{-9}$    &    3.9$\times 10^{-9}$   \\   
X(HCN)                  &     7.0$\times 10^{-9}$    &    2.4$\times 10^{-9}$    &    9.0$\times 10^{-9}$   \\                      
X(CS)                     &     1.3$\times 10^{-8}$    &    4.8$\times 10^{-9}$    &    3.3$\times 10^{-9}$    \\         
X(SO)                     &     1.8$\times 10^{-9}$    &    2.9$\times 10^{-9}$    &    8.8$\times 10^{-10}$   \\                  
X(HCS$^+$)           &     1.4$\times 10^{-10}$   &    3.3$\times 10^{-11}$   &    3.6$\times 10^{-11}$   \\ 
X(N$_2$H$^+$)     &     7.7$\times 10^{-10}$    &    1.0$\times 10^{-9}$    &    9.6$\times 10^{-10}$    \\
\hline
R$_{\rm TD}$(CO)                   &     1.7        &        2.6         &         1.2        \\
R$_{\rm TD}$(HCO$^+$)       &     1.3        &        1.9         &         3.3         \\
R$_{\rm TD}$(HCN)                &     0.1        &        0.4         &         0.1         \\
R$_{\rm TD}$(CS)                   &     1.5        &        4.2         &         6.0         \\
R$_{\rm TD}$(SO)                   &     2.2        &        1.4         &         4.5          \\
R$_{\rm TD}$(HCS$^+$)         &     0.3        &        1.5         &         $<$1       \\
R$_{\rm TD}$(N$_2$H$^+$)   &     $<$0.8  &        $<$0.6   &         $<$0.6      \\
\hline \hline
\end{tabular}

\noindent
Notes: Molecular abundances with respect to H$_2$ towards the (0$"$,0$"$) positions of TMC~1-CP, TMC~1-NH3 and TMC~1-C. 
R$_{\rm TD}$(X) is the ratio between the abundance of the species X in the translucent phase (red lines in Fig.~\ref{mol1} and \ref{mol2}) over the abundance towards each cut extinction peak [i.e., offset (0$"$,0$"$)]. 
\end{table}

\section{Dense phase}
In the following, we carry out a phenomenological analysis of the chemical changes observed 
across the cuts in the dense phase. 
The abundance of most molecules decreases with the visual extinction from A$_V$ = 10 to $\sim$20 mag. 
In Table~6 we show the estimated molecular abundances and the values of R$_{\rm TD}$(X), defined as the ratio between the abundance 
of the species X in the translucent phase (red lines in Fig.~\ref{mol1} and \ref{mol2}) over the abundance towards 
each cut extinction peak [offset (0$"$,0$"$)].  Values of R$_{\rm TD}$(CO), R$_{\rm TD}$(HCO$^+$), R$_{\rm TD}$(CS) 
and R$_{\rm TD}$(SO) $>$1 are found in the three studied cuts. We cannot derive any conclusion, however, 
on HCS$^+$ with very few detections in the translucent phase and 
an abundance a factor of $\sim$4 larger towards TMC~1-CP than towards the TMC~1-NH3 and TMC1-C.
The value of R$_{\rm TD}$(HCO$^+$) is very dependent on the CO depletion and
the change in the gas ionization fraction in the higher density core center,
X(e$^-$) $\propto$ $\sqrt{\zeta _{H_2}/n}$.  The variations in the CO, CS and 
SO abundances are better understood as the 
consequence of the freeze-out of S- and O-bearing molecules onto the grain
mantles.  The depletions of these molecules are slightly higher towards TMC~1-C than
towards TMC~1-NH3 and TMC1-CP, suggesting a different density structure
and/or a more evolved chemical state for the former.  

\begin{figure*}
\includegraphics[angle=0,scale=.2]{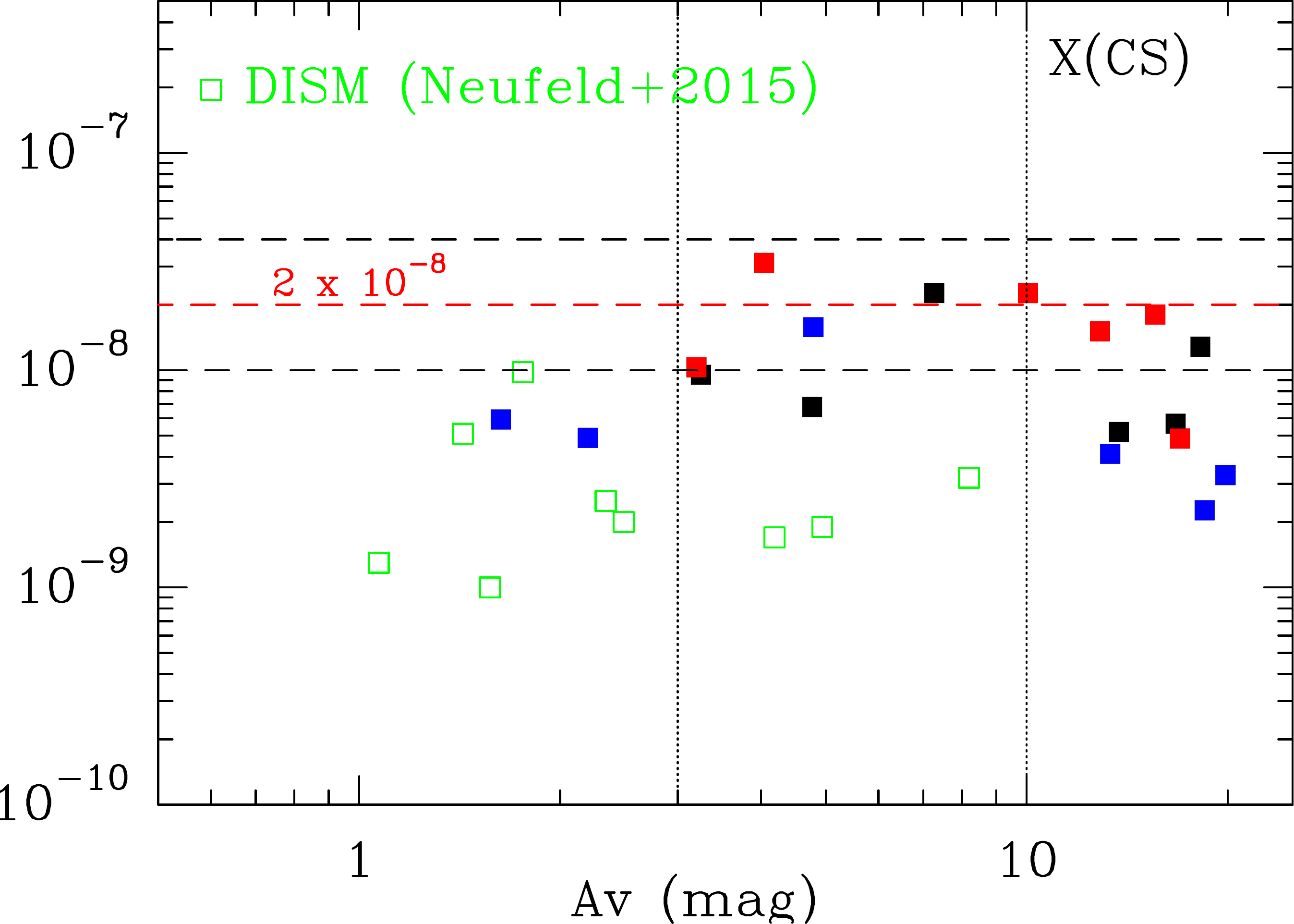}
\includegraphics[angle=0,scale=.2]{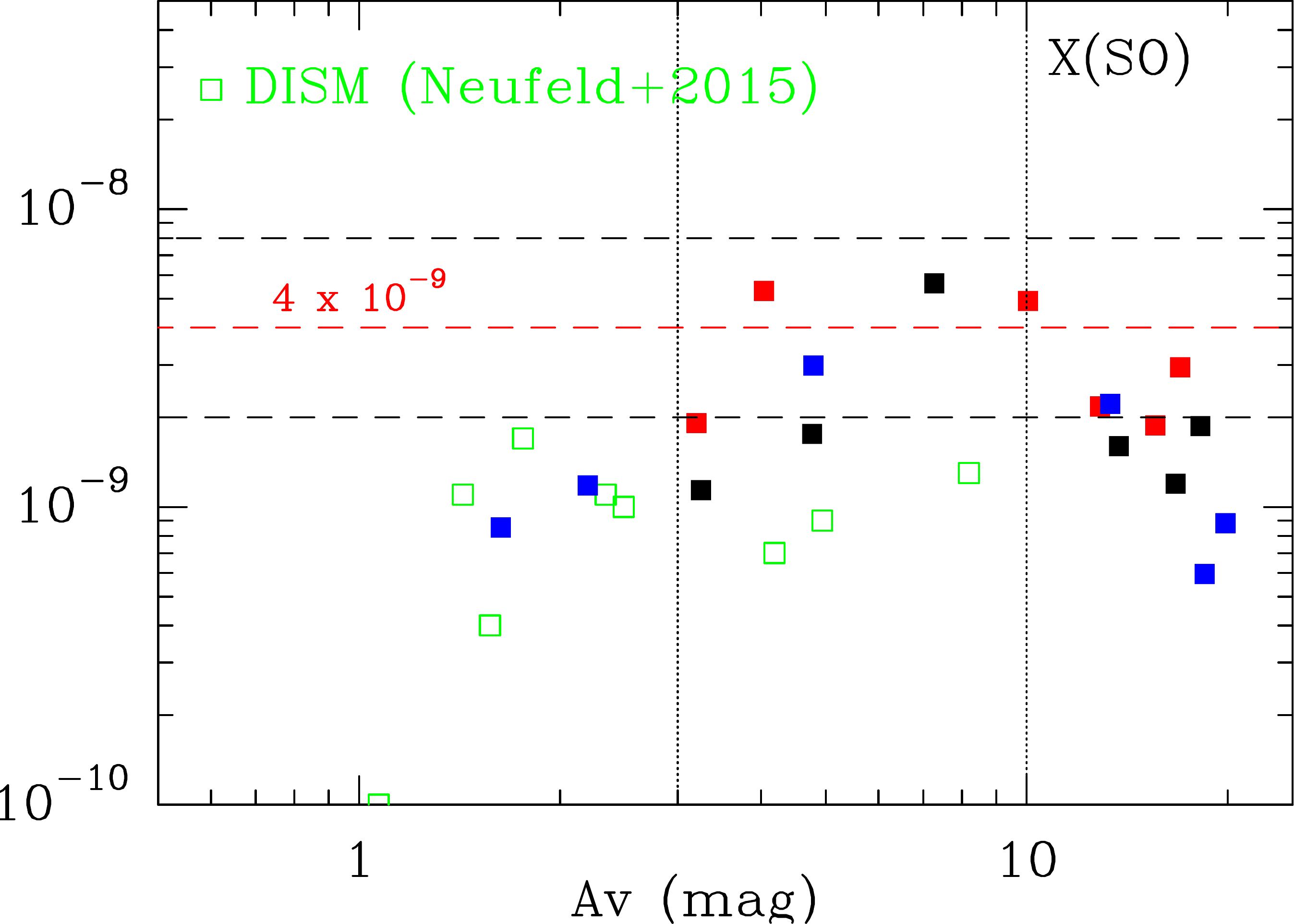}
\includegraphics[angle=0,scale=.2]{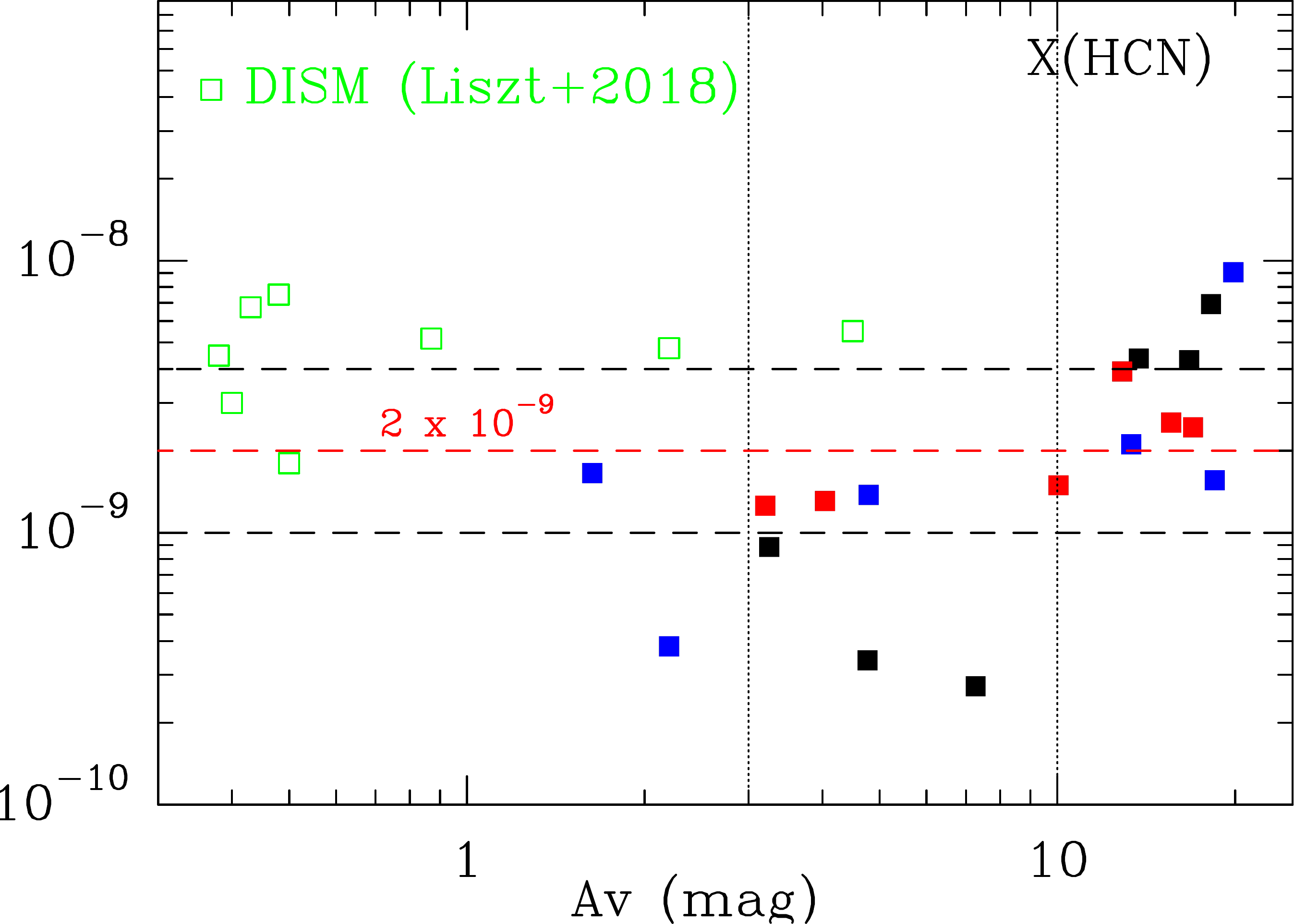}
\caption{Comparison between the abundances in TMC 1 with those observed in the diffuse gas 
by \citet{Neufeld2015} and \citet{Liszt2018}.}
\label{difuso}
\end{figure*}

\begin{table*}
\label{deple}
\begin{tabular}{llllllll}\\
\multicolumn{8}{c}{Table 7 .- Elemental gas phase abundances} \\ \hline \hline
\multicolumn{1}{c}{}     & 
\multicolumn{1}{c}{Solar$^1$}     & 
\multicolumn{1}{c}{DIFF$_{min}$$^2$} & 
\multicolumn{1}{c}{DIFF$_{max}$$^3$} & 
\multicolumn{1}{c}{TRANSLUCENT$^4$}  &  
\multicolumn{1}{c}{TMC~1-CP$^5$}    &
\multicolumn{1}{c}{Orion KL$^6$}     &
\multicolumn{1}{c}{L1157-B1$^7$}   \\ \hline \hline                                                  
C/H  &  2.88$\times$10$^{-4}$ & 2.20$\times$10$^{-4}$ & 1.76$\times$10$^{-4}$ &  {\bf 8.00$\times$10$^{-5}$} &
 9.00$\times$10$^{-5}$ & 1.79$\times$10$^{-4}$ &  1.79$\times$10$^{-4}$ \\
O/H  &  5.75$\times$10$^{-4}$ & 5.50$\times$10$^{-4}$ & 3.34$\times$10$^{-4}$ &  {\bf 8.00$\times$10$^{-5}$} &
 6.40$\times$10$^{-5}$ & 4.45$\times$10$^{-4}$  &  4.45$\times$10$^{-4}$ \\
S/H  &  1.50$\times$10$^{-5}$ & 1.25$\times$10$^{-5}$ & 3.50$\times$10$^{-6}$ &  {\bf 8.00$\times$10$^{-7}$} &
 $<$8.00$\times$10$^{-8}$ & 1.43$\times$10$^{-6}$ & 6.00$\times$10$^{-7}$ \\
 \hline \hline
\end{tabular}

\noindent
Ref:$^1$ Lodders et al. (2003); $^2$ Minimum depletion case of Jenkins (2009); $^3$Maximum depletion case of
Jenkins (2009); $^4$ This work;\\
$^5$ Ag\'undez \& Wakelam (2013); Vidal et al. (2017); $^6$ Esplugues et al. (2014); 
$^7$ Holdship et al. (2016)

\end{table*}

\section{Gas chemical composition from the diffuse to the translucent phase}
In general we refer to 
the gas with densities of n$_H$ $<$ 100 cm$^{-3}$ and T$_k$$\sim$100 K as diffuse gas.
In this phase, the gas is partially  atomic and CO is not a good tracer of the total mass of molecular
gas. The molecular content of the diffuse gas has been determined by a series of
studies based on the molecular absorption lines at millimeter wavelengths which revealed 
a surprisingly rich chemistry (see \citealp{Liszt2018} and references therein).  Translucent clouds
are characterized by n$_H$ $\sim$ a few 1000 cm$^{-3}$,  T$_k$$\sim$20$-$30 K,
the gas is mostly in molecular form and CO is a good mass tracer. The higher densities
of this phase permit the detection of low-excitation molecular emission lines. Although difficult, 
the comparison of the chemical composition
of the diffuse and translucent phases might provide important clues for the understanding of
the chemical evolution of the gas in the interstellar medium. All the species studied in this paper, 
except N$_2$H$^+$, have been detected in the diffuse gas. Interestingly,  we have only one 
detection of N$_2$H$^+$  at A$_V$$\sim$5 mag with N(N$_2$H$^+$)/N(HCO$^+$) $\sim$0.02. 
\citet{Liszt2001} measured N(N$_2$H$^+$)/N(HCO$^+$) $<$ 0.002 
towards 3C111. This quasar is actually seen through a small hole (region of lower than average extinction) in an outlying cloud
in the Taurus cloud complex \citep{Lucas1998} and hence A$_V$$\sim$ 5 mag can be considered as a threshold for the N$_2$H$^+$ detection in Taurus. 

In Fig.~\ref{difuso} we show  the comparison of the abundances with respect to H$_2$ of the studied molecules with those from
our survey. The data towards 3C111 are indicated.  There is a large dispersion in the plot of the molecular abundances
as a function of the visual extinction. We recall that diffuse clouds are not only characterized by low values of the
visual extinction but also for low hydrogen densities. In addition, in the diffuse gas, with several clouds along the line of sight,
the visual extinction is not necessarily related to the local UV field.
One can find a better correlation if one considers the abundances versus the local density and assumes that
in the diffuse gas the local density is around 50-100 cm$^{-3}$. For HCO$^+$, SO and CS we find a trend with their abundances
increasing with density from the diffuse to the translucent phase. All these molecules present their peak abundances
in the translucent region. HCN might be the one exception to this rule with lower abundances in the translucent phase than in the diffuse gas.
As discussed above, the HCN column densities might be severely underestimated in the translucent cloud. 
It is also interesting the case of HCS$^+$ which was detected in the diffuse medium by \citet{Lucas2002} with
an abundance ratio X(CS)/X(HCS$^+$)$\sim$13$\pm$1 that is $\sim$40 times lower than the value of X(CS)/X(HCS$^+$)$\sim$400 we have measured in the translucent cloud. This unveils a different formation path of CS in diffuse (dissociative recombination of
HCS$^+$) and translucent (SO + C$\rightarrow$CS + O) clouds.

\section{Elemental depletions and grain growth }

The depletion of an element X in the ISM is defined in terms of its reduction
factor below the expected abundance relative to that of hydrogen if all of the atoms were in the
gas phase,
\begin{equation}
[Xgas/H] = log{N(X)/N(H)} - log(X/H)_\odot
\end{equation}
In this expression, N(X) is the column density of
element X and N(H) represents the column density of hydrogen in both atomic and molecular
form, i.e., N(H I) + 2N(H$_2$). The missing atoms of element X are presumed to be locked up in
solids within dust grains or in the icy mantle.
In the diffuse gas, atomic absorption lines can be used to determine abundances by comparison
with the atomic and molecular hydrogen column densities measured through
Lyman alpha and Lyman-Werner transitions. \citet{Jenkins2009} presents a comprehensive
study of the elemental depletions in diffuse clouds. 
In general, depletions increase with the average density along the line of sight. 
However, depletions are observed to vary from one line of sight to another.
\citet{Savage1996} interpreted these variations in terms of 
averages of warm (presumably low density) gas and cool (denser) gas. 
In his review, \citet{Jenkins2009} distinguishes between two cases, "minimum"
and "maximum" depletion to characterize the range
of these variations in diffuse clouds.

In Table~7 we 
compare our estimates in the translucent part of TMC1 with the "minimum"
and "maximum" depletion cases, in diffuse clouds. Our value of C/H
is consistent within a factor of 2  with the $"$maximum$"$ depletion diffuse case. 
A significant difference is found, however, in  the O and S depletions which are $\sim$4 times larger 
in the translucent phase than in the "maximum" depletion case. 
Although we observe a smooth increase
in the C and O depletions with visual extinction in the translucent phase, 
the C/O ratio remains quite constant ($\sim$1).
The  scatter in the measured X(CS)/X(SO) values can hinder 
a smooth variation in the derived C/O value but an almost constant C/O ratio can be understood
if the freeze-out of CO is the main process that changes the grain composition in this region.
Regarding sulfur, we measure a depletion factor of $\sim$7-40 in the translucent cloud.
Although some authors like \citet{Jenkins2009} casts doubts on this interpretation,
it is widely accepted that sulfur is not depleted in diffuse clouds (see also
\citealp{Neufeld2015}). Adopting this scenario,
sulfur atoms (or ions) should be massively incorporated to dust grains from the HI/H$_2$ 
(A$_V$$\sim$1) to the C$^+$/C/CO transition phases to explain a 
depletion factor of $\sim$7-40 in the translucent medium. 

\begin{figure*}
\includegraphics[angle=0,scale=.55]{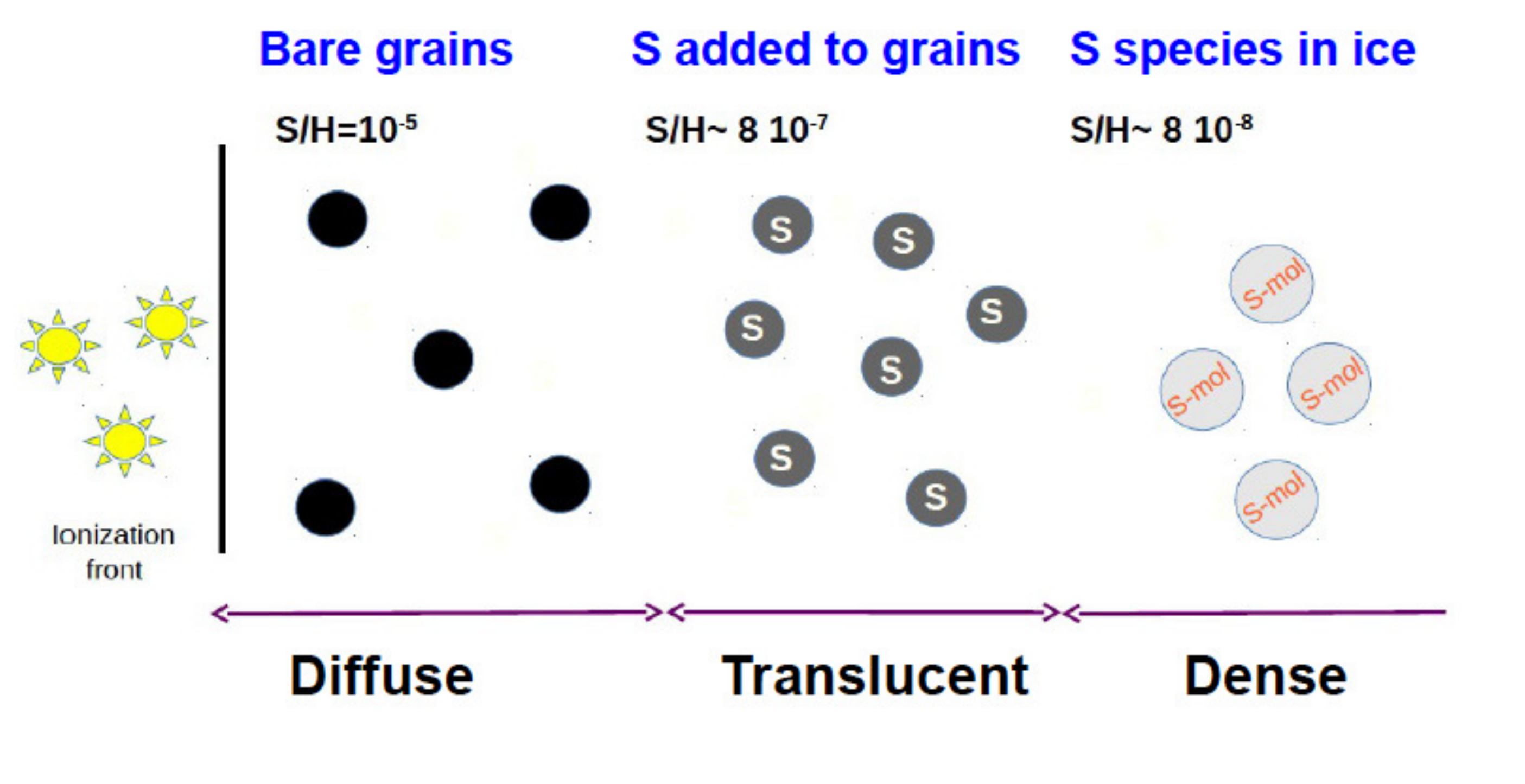}
\caption{Scheme of the proposed scenario of the sulfur gas-phase/solid state evolution in TMC~1 (courtesy of S. Cazaux).} 
\label{cart}
\end{figure*}

In Table~7 we also compare the depletions estimated in the translucent part of 
TMC~1 with the chemical composition towards TMC~1-CP. 
We have estimated the C and O depletions from the CO abundance reported by \citet{Agundez2013} and
assumed C/O$\sim$1.  The S depletion is based on recent results of \citet{Vidal2017}.
While C and O depletions agree within a factor of 2 with the values in the translucent cloud,  
sulfur depletion needs to be increased by at least a 
factor 10 (depletion of  $\sim$200 compared to solar) to account for the observed
abundances of S-bearing molecules.  

Following these findings, we propose that two strong S depletion events should occur across the cloud. The first 
occurs in the transition from the diffuse to the translucent phase. In this transition, $\sim$90\% of  
the sulfur is incorporated into dust grains while $\sim$10\% is hidden as 
atomic sulfur in the gas phase. The second strong depletion  occurs in the dense gas where a thick ice 
mantle is formed on the grain surfaces (see the illustrative scheme
in Fig.~\ref{cart}). The exact composition of the icy mantle has not established yet.
For large values of C/O and early times, atomic sulfur would remain as the
main sulfur reservoir in the dense gas.  These S atoms would become adsorbed on the 
icy surface, and would react with hydrogen atoms to produce solid HS and H$_2$S in ices. These latest 
reservoirs are supported by observations from the comet 67P with Rosetta showing that H$_2$S 
is the most important S-bearing species in cometary ices \citep{Calmonte2016}.
For low C/O ratios and close to the steady state, chemical models predict that most sulfur is in molecular 
form as SO and SO$_2$. These molecules are rapidly  frozen onto dust grains at high densities and temperatures 
below $\sim$50 K (see, e.g., \citealp{Pacheco2016}), trapping the sulfur in the solid phase.
Observationally, OCS is the only S-bearing molecule unambiguously detected in ice
mantles because of its large band strength in the infrared \citep{Palumbo1995} and, tentatively, SO$_2$
\citep{Boogert1997}. H$_2$S has not been detected in interstellar ices through
infrared absorption experiments \citep{Jimenez-Escobar2011}.

Regardless of the exact chemical composition, the sulfur budget in the ice is expected to return to 
the gas phase in hot cores and bipolar outflows. 
The study of the sulfur chemistry in these environments can therefore provide some clues on the fate of sulfur. 
In Table~7, we show the values of sulfur depletion derived towards the hot core Orion KL  \citep{Esplugues2014}
and the shocked region L1157-B1 \citep{Holdship2016}. Interestingly, the sulfur depletion in these two
sources is $\sim$10-30 relative to the solar value, i.e.,  similar to the value we measured in the translucent
phase. A similar value of sulfur depletion was measured in bipolar outflows by \citet{Anderson2013} 
using observations of the infrared space telescope Spitzer. This suggests that the fraction of sulfur 
incorporated onto the grains in the diffuse-translucent transition ($\sim$90\%) is not released to the gas 
phase when the icy mantles are destroyed.  This has important implications for star and planet formation 
studies. During the formation of a low-mass star,  the grain cores are not destroyed 
except in the bow shocks formed at the tip of high-velocity jets. 
If our hypothesis is correct, 90\% of the S atoms would remain locked in the grains
in the inner regions of protoplanetary disks where planet formation occurs.

\section{Summary and conclusions}
This paper is based on the Gas phase Elemental abundances in molecular
CloudS (GEMS, PI: A. Fuente) of the prototypical dark cloud TMC~1.
The Taurus molecular cloud (TMC) is one of the closest, low-mass star-forming regions at 140 pc. 
In this paper we investigate the chemistry to derive the elemental gas-phase abundances and
the gas ionization fraction in the translucent part (A$_V <$ 10 mag) of this molecular
cloud. The chemistry in this transition from the diffuse to the dense gas  
determines the initial conditions for the formation of the dense contracting cores. 

\begin{itemize}
\item   We use millimeter observations of a selected sample of species carried out with the IRAM 30m telescope (3\,mm and 2\,mm) and the 40m Yebes 
 telescope (1.3\,cm and 7\,mm)) to determine the fractional abundances of CO, HCO$^+$, HCN, CS, SO, HCS$^+$, and N$_2$H$^+$ in positions  
 along three cuts intersecting the main filament  at  positions TMC~1-CP, TMC~1-NH3, and TMC~1-C over which the visual extinction varies
 between peak values of A$_V$$\sim$20 mag and 3 mag. 

\item None of the studied molecules presents constant abundance across the studied cuts. According to their variations 
with visual extinction, we can differentiate three groups. The first group is formed by the abundant molecule $^{13}$CO.
This molecule reaches its peak value at A$_V$$\sim$ 3 mag and then progressively decreases with visual extinction.
The second group is formed by HCO$^+$, CS, and SO; the abundances of these molecules increases with visual
extinction until A$_V$$\sim$ 5 mag where they present a narrow peak and then progressively decreases towards the
extinction peak. In the third group, the abundance of the N-bearing molecules HCN and N$_2$H$^+$ increases from A$_V$$\sim$3~mag
until the extinctions peaks at  A$_V$$\sim$20~mag.
 
 \item By comparison of the molecular abundances with the Meudon PDR code,  we derive the C, O, and S elemental depletions, and hence 
 the gas ionization degree as a function of the visual extinction at each position. Our data show that even at A$_V$$\sim$ 3$-$4 mag where the  transition C$^+$/C/CO occurs, significant depletions of C, O, and S  are found. In fact, C/H varies between  $\sim$8 10$^{-5}$ and $\sim$4 $\times$ 10$^{-5}$ in the translucent cloud (3 $<$ A$_V$ $<$ 10 mag). Moreover,  the C/O ratio is $\sim$ 0.8$-$1, suggesting that the O is preferentially depleted in the diffuse phase (A$_V$ $<$ 3 mag). Regarding sulfur, we  estimate S/H $\sim$ 0.4 $-$ 2.2  $\times$10$^{-6}$ in this moderately dense region. 

 \item The detailed modeling of the chemistry in the translucent phase and our estimate of the elemental abundances allow us to constrain the value of  $\zeta_{\rm H_2}$ to $\sim$ 0.5$-$1.8$\times$10$^{-16}$ s$^{-1}$. This value is slightly lower (a factor of $\sim$3) than that 
 derived by \citet{Neufeld2017}, $\zeta_{\rm H_2}$ = (5.3$\pm$1.1)$\times$10$^{-16}$ s$^{-1}$, in the diffuse medium.
 
\end{itemize}

Based on our results, we propose that the freeze out of CO
is the main process that changes the grain composition in the translucent part of the 
cloud producing a progressive depletion of C and O from A$_V$$\sim$ 3 mag to
 A$_V$$\sim$ 10. Regarding sulfur, we measure a constant depletion of $\sim$7$-$40 across 
 the translucent cloud. This suggests that
sulfur atoms (or ions) would have been massively incorporated onto dust grains from the HI/H$_2$ 
(A$_V$$\sim$1) to the C$^+$/C/CO transition to reach a 
depletion of $\sim$7-40 in the translucent medium. In order to account for the chemical
composition in the TMC~1-CP core, a second strong S depletion should occur
in the dense cloud. Interestingly, the S atoms incorporated into grains during the diffuse and
translucent phase are not returned back to the gas phase during the formation of a low-mass star.

\begin{acknowledgements}
We thank the Spanish MINECO for funding support from AYA2016-75066-C2-1/2-P,  and the ERC under ERC-2013-SyG, G. A. 610256 NANOCOSMOS. JM acknowledges the support of ERC-2015-STG No. 679852 RADFEEDBACK. SPTM acknowledges to the European Union's Horizon 2020 research and innovation program for funding support given under grant agreement No~639459 (PROMISE). RMD acknowledges support provided by an award from the Simons Foundation (SCOL\#321183, KO). GMC acknowedges funding support fromAYA2017-85322-R. MT acknowledges partial support from project AYA2016-79006-P."
\end{acknowledgements}


\appendix

\section{New calculations of the SO + OH $\rightarrow$ SO$_2$ + H reaction rate}

The potential energy surface (PES) of the ground electronic state of HSO$_2$
system have been developed by \cite{Ballester-Varandas:05} by fitting
very accurate {\it ab initio} calculations. According to this PES,
the SO($^3\Sigma$) + OH($^2\Pi$) $\rightarrow$ SO$_2$+ H
reaction is exothermic by $\approx$ 1.3 eV, with two deep wells, of $\approx$ 3 eV for HOSO
and of $\approx$ 2 eV for HSO$_2$. There is a barrier between the two wells, of energy very close to the 
SO + OH asymptote. The HOSO is directly connected to the   SO + OH with no barrier, with the attractive
dipole-dipole long-range interaction.

There have been several quasi-classical trajectory (QCT) calculations
of the  SO + OH $\rightarrow$ SO$_2$+ H rate \citep{Ballester-Varandas:07,Ballester-etal:10,Pires-etal:14}
finding that this reaction presents a capture-like behavior due to the dipole-dipole interaction
and a very good agreement with the available experimental rates \citep{Blitz-etal:00,Jourdain-etal:79,Fair-Thrush:69}.
All these experimental and theoretical results were obtained at temperatures above 200~K,
and in this work we extend the simulations to lower temperatures, down to 10~K, of interstellar
interest.

\begin{figure}[h]
\vspace*{+1.5cm}
\hspace*{-1.40cm}
\includegraphics[scale=0.90,angle=0]{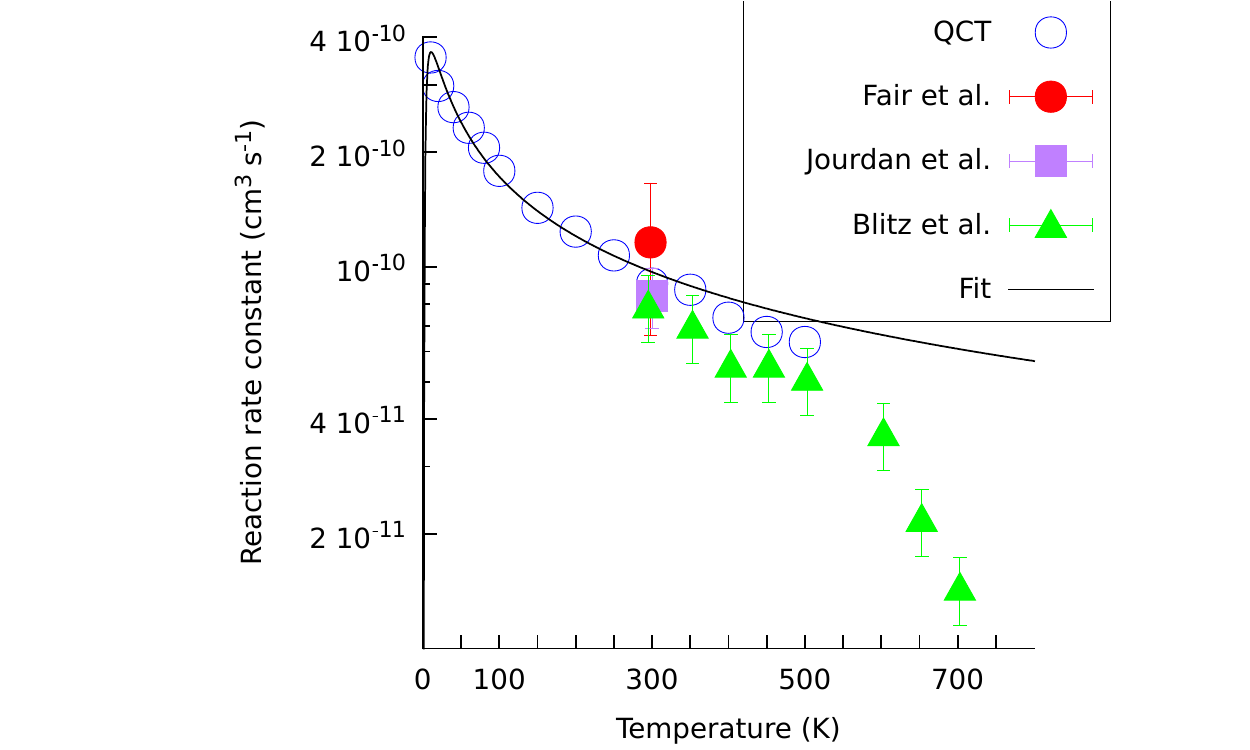}
   \caption{\label{rates-fig}
State-selected reaction rates for SO$(v=0)$ + OH(v=0) $\rightarrow$ SO$_2$ + H collisions
as a function of  temperature obtained in this work (QCT), and compared with the experimental
results of previous works \citep{Blitz-etal:00,Jourdain-etal:79,Fair-Thrush:69}.
The calculated rate has been fitted to the expression $ K(T)=a (T/300)^b  e^{-c/T}$,
with a=1.24211 10$^{-10}$ cm$^3$/s,b=$-$0.56049  and
c=6.58356 K. This expression is adequate in the 10$-$300~K interval.
}
\end{figure}

QCT calculations have been performed at temperatures in the 10-500K interval,
using the miQCT code \citep{Dorta-Urra-etal:15,Zanchet-etal:18}.
For each temperature batches of 50000 trajectories were run, starting at a distance
between SO and OH center-of-mass of 125 Bohr. The initial impact parameter, $b$,
was randomly chosen between 0-90 Bohrs according to a $b^2$ distribution. The initial
translational and rotational energy of the two reactants was randomly chosen according
to a Boltzmann distribution, while the vibrations of the two reactants was described using a adiabatic switching
method \citep{Grozdanov-Solovev:82,Johnson:87,Qu-Bowman:16,Nagy-Lendvay:17} corresponding to the ground
vibrational state of the two reagents. The reactivity of this
reaction does not depend on the initial vibrational excitation
of the reactants\citep{Ballester-Varandas:07,Ballester-etal:10,Pires-etal:14}. For this reason, we
can consider that the vibrational state selected rate calculated here is essentially
the thermal rate constant. The reactive rate constant is calculated as
\begin{eqnarray}
k_{vj}(T) = g_e(T) \,\sqrt{ {8k_B T\over \pi \mu} }\, {N_r\over N_t}\, \pi b_{max}^2,
\end{eqnarray}
where $N_r$ denotes the number of reactive trajectories and $b_{max}$ is the maximum
impact parameter for which reaction takes place. Finally, $g_e$ is the electronic
partition function 
\begin{eqnarray}
g_e(T)= \left\lbrace
          3\left( 1+ e^{-209/T}\right)
        \right\rbrace^{-1},
\end{eqnarray}
where the factor 3 arises because only one combination of total spin 1/2 is reactive,
when combinaing the triplet state of SO and the double state of OH, and that
the OH($^2\Pi$) splits into two spin-orbit states ($^2\Pi_{3/2}$ and $^2\Pi_{1/2}$, with
an energy difference of 209 K. The results obtained are shown in Fig.~\ref{rates-fig}
and compared with the experimental results available.

The QCT results made in this work are in rather good agreement with the experimental data
in the interval 300-500~K. The rate increases by a factor of 3-4 when decreasing the temperature
from 300 to 10~K. This may have some consequences in astrophysical objects at low temperatures.
It should be noted that the fit presented in he caption of Fig.~\ref{rates-fig} is only adequate for  temperatures between 10 and 300~K.

\section{Tables and Figures}

\begin{table*}
\label{setup}
\begin{tabular}{llllll}\\
\multicolumn{6}{l}{Table B.1- Telescope parameters} \\ \hline \hline
\multicolumn{1}{c}{Telescope}  &  \multicolumn{1}{c}{Setup}       & \multicolumn{1}{c}{Freq. band} &  
\multicolumn{1}{c}{HPBW($"$)}  & \multicolumn{1}{c}{$F_{eff}$}  & \multicolumn{1}{c}{$B_{eff}$} \\ \hline \hline

IRAM 30m &  Setup 1   &   L106  &  24  &  0.95  & 0.80  \\ \hline

         &  Setup 2   &   L89   &  29  &  0.95  & 0.81  \\
         &            &   L147  &  16  &  0.93  & 0.74  \\ \hline
         
         &  Setup 3   &   L101  &  25  &  0.95  & 0.80  \\
         &            &   L138  &  17  &  0.93  & 0.74  \\ \hline
         
         &  Setup 4   &   L92   &  17  &  0.95  &  0.81 \\
         &                 &   L168  &  14  &  0.93  &  0.74 \\ \hline 
         
Yebes 40m  & Setup 0   &   L23000   &  84  &  0.93  &  0.70 \\
                   &                 &   L44750   &  42  &  0.90  &  0.49 \\                   
\hline \hline
\end{tabular}
\end{table*}

\begin{table*}
\label{species}
\begin{tabular}{llrlll} \\

\multicolumn{6}{l}{Table B.2 .- Spectral setups} \\ \hline \hline
\multicolumn{2}{c}{Line}     & \multicolumn{1}{c}{Freq.(MHz)} &  \multicolumn{1}{c}{E$_u$(K)} &
 \multicolumn{1}{c}{A$_{ul}$(s$^{-1}$)}  &  \multicolumn{1}{c}{g$_u$}  \\  \hline \hline
\multicolumn{6}{c}{L89}  \\ 
HCS$^+$     & 2$\rightarrow$1        & 85347.87  &  6.1  &  1.110 10$^{-5}$  &  5    \\
HCN         & 1$\rightarrow$0        & 88631.85  &  4.3  &  2.406 10$^{-5}$  &  3    \\
H$^{13}$CN  & 1$\rightarrow$0        & 86340.18  &  4.1  &  2.224 10$^{-5}$  &  3    \\
HC$^{15}$N  & 1$\rightarrow$0        & 86054.97  &  4.1  &  2.202 10$^{-5}$  &  3    \\ 
HCO$^+$     & 1$\rightarrow$0        & 89188.53  &  4.3  &  4.234 10$^{-5}$  &  3    \\
H$^{13}$CO$^+$ & 1$\rightarrow$0     & 86754.29  &  4.2  &  3.897 10$^{-5}$  &  3    \\
HC$^{18}$O$^+$ & 1$\rightarrow$0     & 85162.22  &  4.1  &  3.686 10$^{-5}$  &  3    \\
HNC         & 1$\rightarrow$0        & 90663.56  &  4.4  &  2.690 10$^{-5}$  &  3    \\
OCS         & 7$\rightarrow$6        & 85139.10    &  16.3  &  1.715 10$^{-6}$ & 15   \\ 
SO          & 2$_2$$\rightarrow$1$_1$  & 86093.96  &  19.3  &  5.250 10$^{-6}$ &  5   \\ \hline

\multicolumn{6}{c}{L92} \\
$^{13}$CS  &  2$\rightarrow$1       &  92494.27  &  6.7  &  1.412 10$^{-5}$  &  5   \\ 
C$^{34}$S  &  2$\rightarrow$1       &  96412.95  &  6.9  &  1.600 10$^{-5}$  &  5   \\
CH$_3$OH   &  2,-1$\rightarrow$1,-1 &  96739.36  &  4.6  &  2.558 10$^{-5}$  &  5   \\ 
CH$_3$OH   &  2,1$\rightarrow$1,1   &  96741.37  &  7.0  &  3.408 10$^{-6}$  &  5  \\ \hline 
   
\multicolumn{6}{c}{L101} \\
CS        &   2$\rightarrow$1           &    97980.95  &   7.1  &  1.679 10$^{-5}$ &  5  \\  
SO        &   2$_3$$\rightarrow$1$_2$   &    99299.89  &   9.2  &  1.125 10$^{-5}$ &  7  \\  
$^{34}$SO  &  2$_3$$\rightarrow$1$_2$   &    97715.40  &   9.1  &  1.073 10$^{-5}$ &  7  \\   
H$_2$CS    &  3(1,2)$\rightarrow$2(1,2) &    101477.81  &  8.1  &  1.260 10$^{-5}$ &  7  \\  \hline

\multicolumn{6}{c}{L106} \\
$^{13}$CO   &  1$\rightarrow$0   &   110201.35  &   5.3   &  6.336 10$^{-8}$ &  3 \\   
C$^{18}$O  &  1$\rightarrow$0   &   109782.17  &   5.3   &  6.263 10$^{-8}$ &  3 \\   
N$_2$H$^+$ &  1$\rightarrow$0   &   93173.77   &   4.5   &  3.628 10$^{-5}$ &  3  \\ 
SO         & 3$_2$$\rightarrow$2$_1$  &  109252.18   &  21.1 & 1.080 10$^{-5}$ &  5  \\
$^{34}$SO  & 3$_2$$\rightarrow$2$_1$  &  106743.37   &  20.9 & 1.007 10$^{-5}$ &  5  \\  
NH$_2$D    & 1(1,1)$\rightarrow$1(0,1) & 110153.59   &  21.3 & 5.501 10$^{-6}$ &  9  \\  
CH$_3$OH   &  0,0$\rightarrow$1,-1     & 108893.94   &  5.2  & 1.471 10$^{-5}$ &  3  \\  \hline

\multicolumn{6}{c}{L138} \\
$^{13}$CS &  3$\rightarrow$2            &   138739.26  &  13.3  &  5.107 10$^{-5}$  &  7  \\  
SO        &  3$_4$$\rightarrow$2$_3$    &   138178.66  &  15.9  &  3.166 10$^{-5}$  &  9  \\
OCS       &  11$\rightarrow$10          &   133785.90  &  38.5  &  6.818 10$^{-6}$  &  23  \\
H$_2$CS   &  4(0,4)$\rightarrow$3(0,3)  &   137371.21  &  16.5  &  3.647 10$^{-5}$  &  9   \\ 
HDCO      &  2(1,1)$\rightarrow$1(1,0)  &   134284.90  &  17.6  &  4.591 10$^{-5}$  &  5   \\ \hline

\multicolumn{6}{c}{L147} \\
CS        &  3$\rightarrow$2   &  146969.03  &  14.1  &  6.071 10$^{-5}$  &  7 \\     
C$^{34}$S &  3$\rightarrow$2   &  144617.10  &  13.9  &  5.784 10$^{-5}$  &  7 \\  \hline

\multicolumn{6}{c}{L168} \\
H$_2$S       &  1(1,0)$\rightarrow$1(0,1)  &  168762.75  &  8.1   &  2.677 10$^{-5}$ &  3  \\    
H$_2$$^{34}$S   &  1(1,0)$\rightarrow$1(0,1)  &  167910.52  &  8.1   &  2.616 10$^{-5}$ &  3  \\  
HCS$^+$         &  4$\rightarrow$3         &  170691.62  &  20.5  &  9.863 10$^{-5}$ &  9  \\ 
HC$^{34}$S$^+$  &  4$\rightarrow$3         &  167927.25  &  20.1  &  7.805 10$^{-5}$ &  9  \\   
SO              &  4$_4$$\rightarrow$3$_3$  &  172181.42 &  33.8  &  5.833 10$^{-5}$ &  9  \\ 
$^{34}$SO       &  4$_4$$\rightarrow$3$_3$  &  168815.11 &  33.4  &  5.498 10$^{-5}$ &  9  \\ \hline

\multicolumn{6}{c}{L23000} \\  
NH$_3$  &  (1,1)a$\rightarrow$(1,1)s  &  23694.49  &  1.1   &  1.712 10$^{-7}$ &  12   \\
NH$_3$  &  (2,2)a$\rightarrow$(2,2)s  &  23722.63  &  42.3  &  2.291 10$^{-7}$ &  20   \\ \hline

\multicolumn{6}{c}{L44500} \\
CS         & 1$\rightarrow$0   &  48990.96  &   2.4  &  1.749 10$^{-6}$ &  3  \\
C$^{34}$S  & 1$\rightarrow$0   &  48206.94  &   2.3  &  1.666 10$^{-6}$ &  3  \\
$^{13}$CS  & 1$\rightarrow$0   &  46247.56  &   2.2  &  1.471 10$^{-6}$ &  3  \\
HCS$^+$    & 1$\rightarrow$0   &  42674.19  &   2.0  &  1.156 10$^{-6}$ &  3  \\ 
OCS        & 4$\rightarrow$3   &  48651.60  &   5.8  &  3.047 10$^{-7}$ &  9  \\   \hline \hline

\end{tabular}
\end{table*}
 
\begin{figure*}
\includegraphics[angle=0,scale=.9]{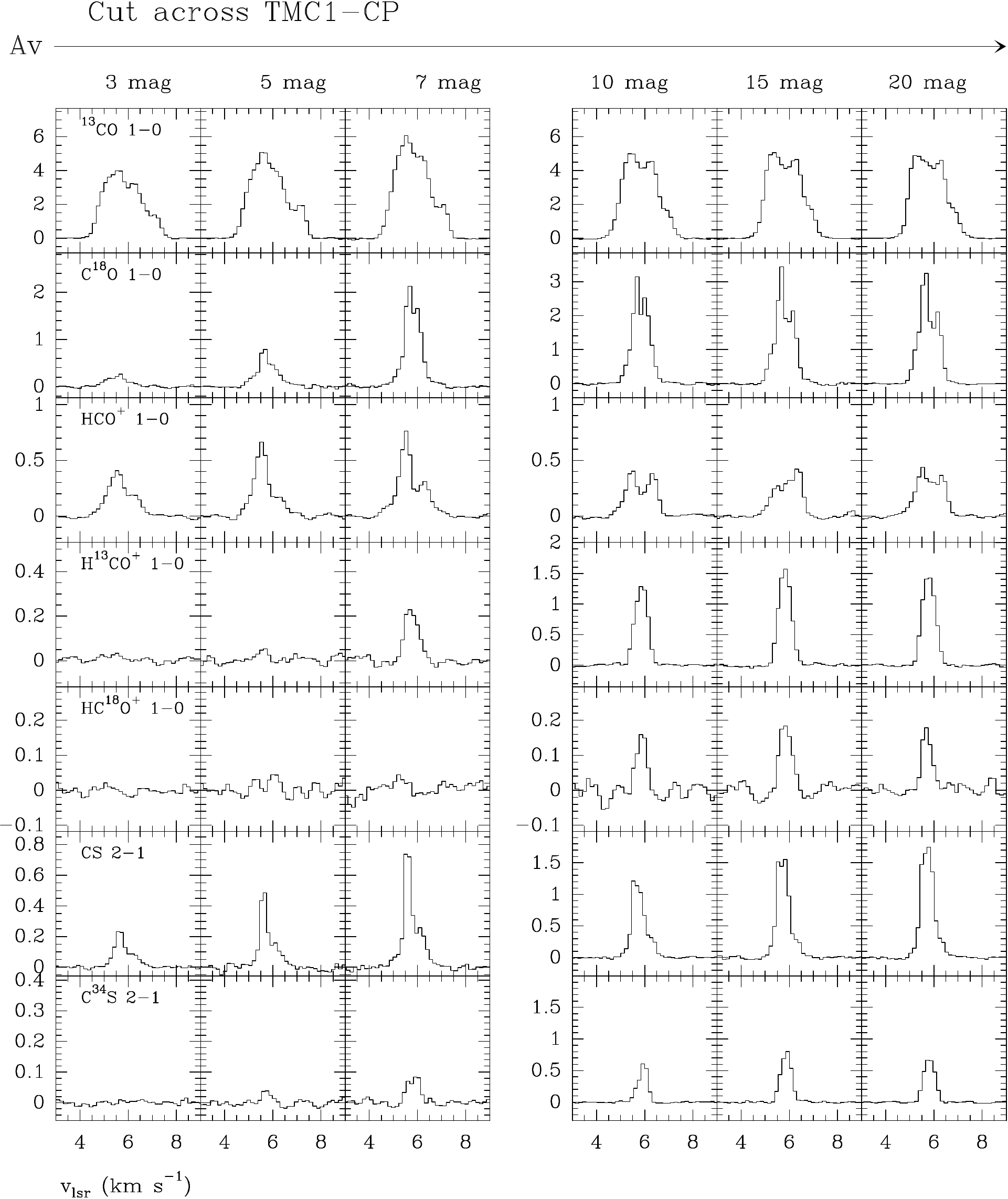}
\caption{Selected sample of spectra as observed with the 30m telescope towards the TMC~1-CP cut (in T$_{MB}$).}
\label{cp-espec1}
\vspace{-0.1cm}
\end{figure*}

\begin{figure*}
\includegraphics[angle=0,scale=.9]{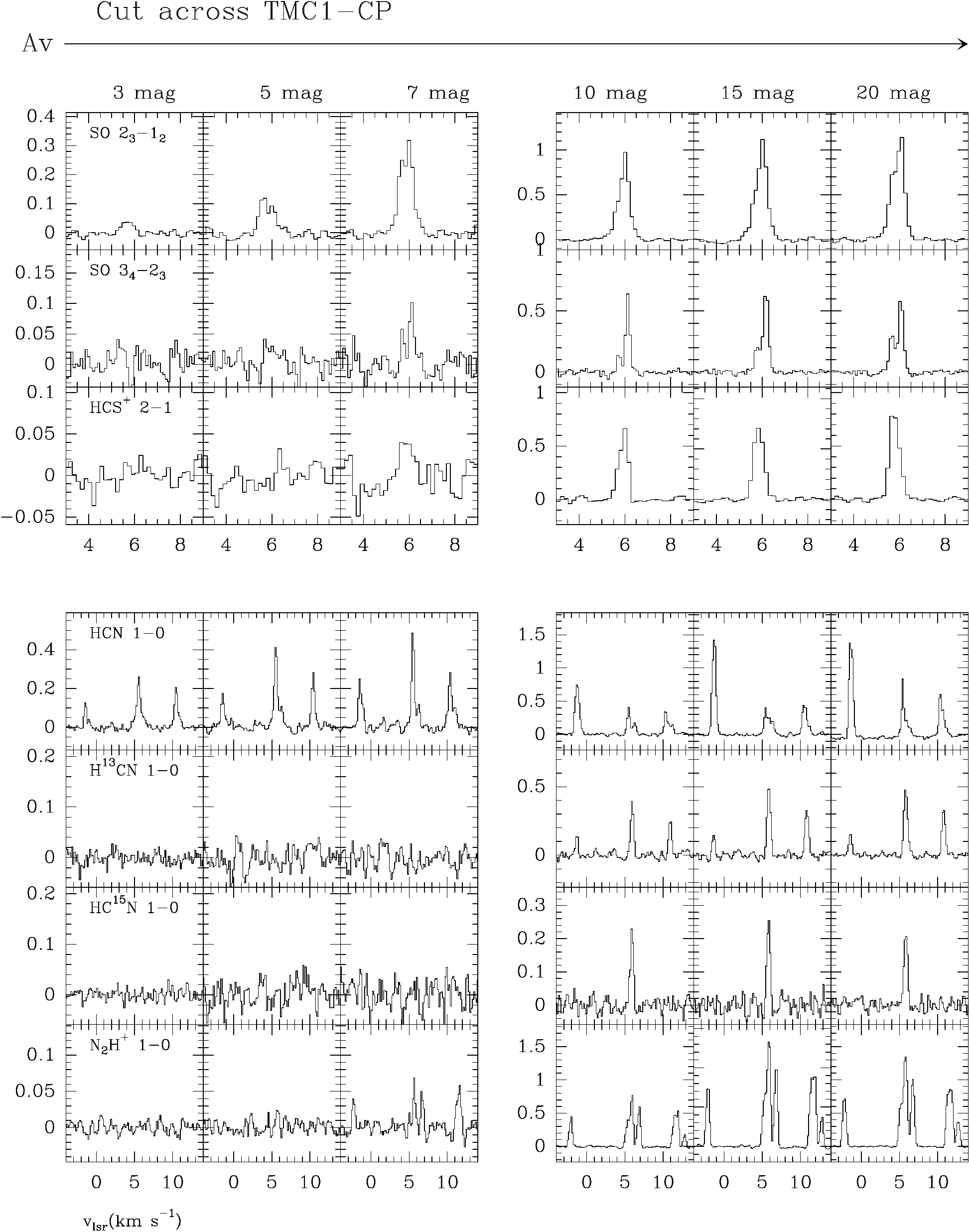}
\caption{The same as Fig. B.1}
\label{cp-espec2}
\vspace{-0.1cm}
\end{figure*}
 
\begin{figure*}
\includegraphics[angle=0,scale=.9]{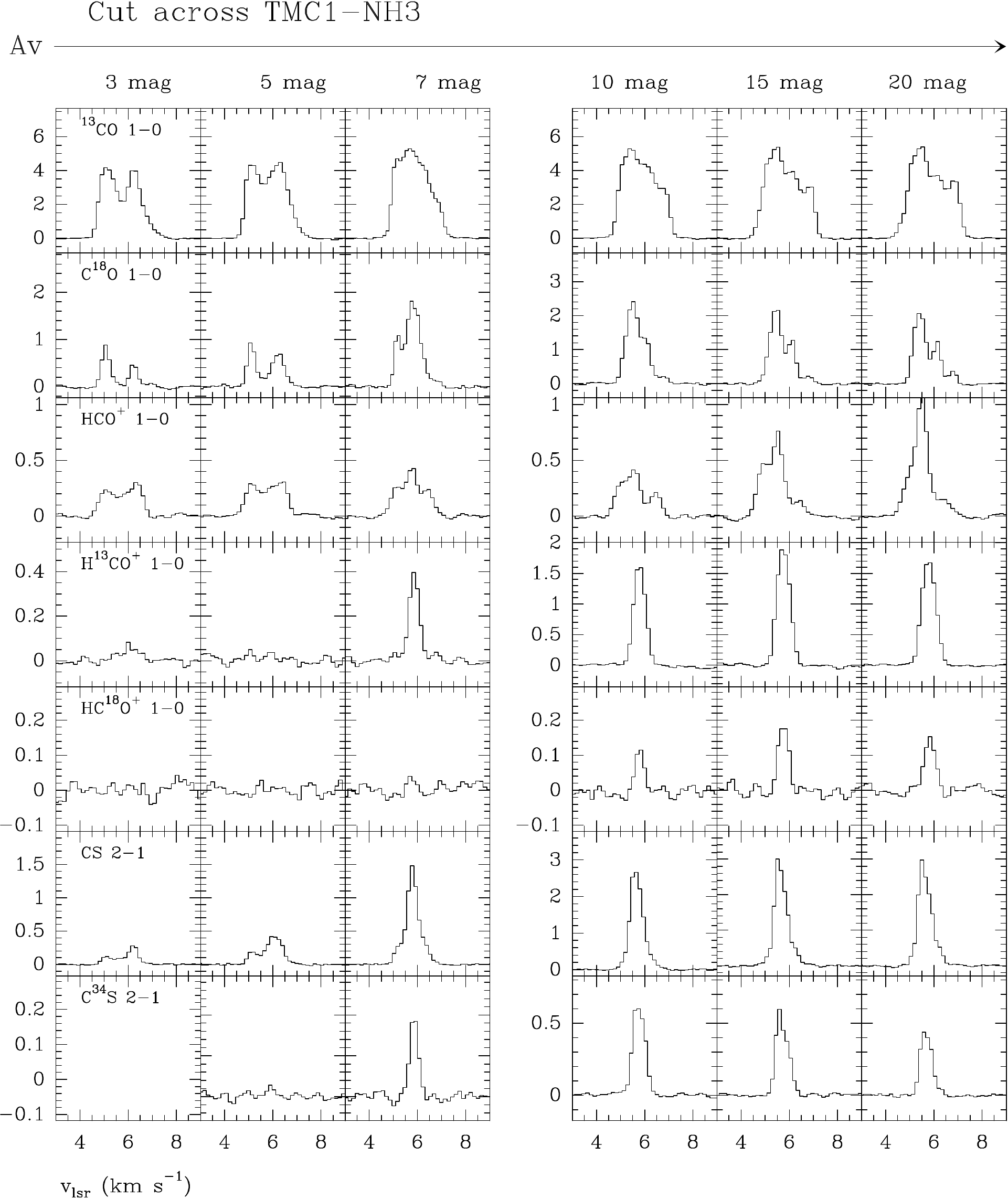}
\caption{Selected sample of spectra as observed with the 30m telescope towards the TMC~1-NH3 cut (in T$_{MB}$).}
\label{nh3-espec1}
\vspace{-0.1cm}
\end{figure*}

\begin{figure*}
\includegraphics[angle=0,scale=.9]{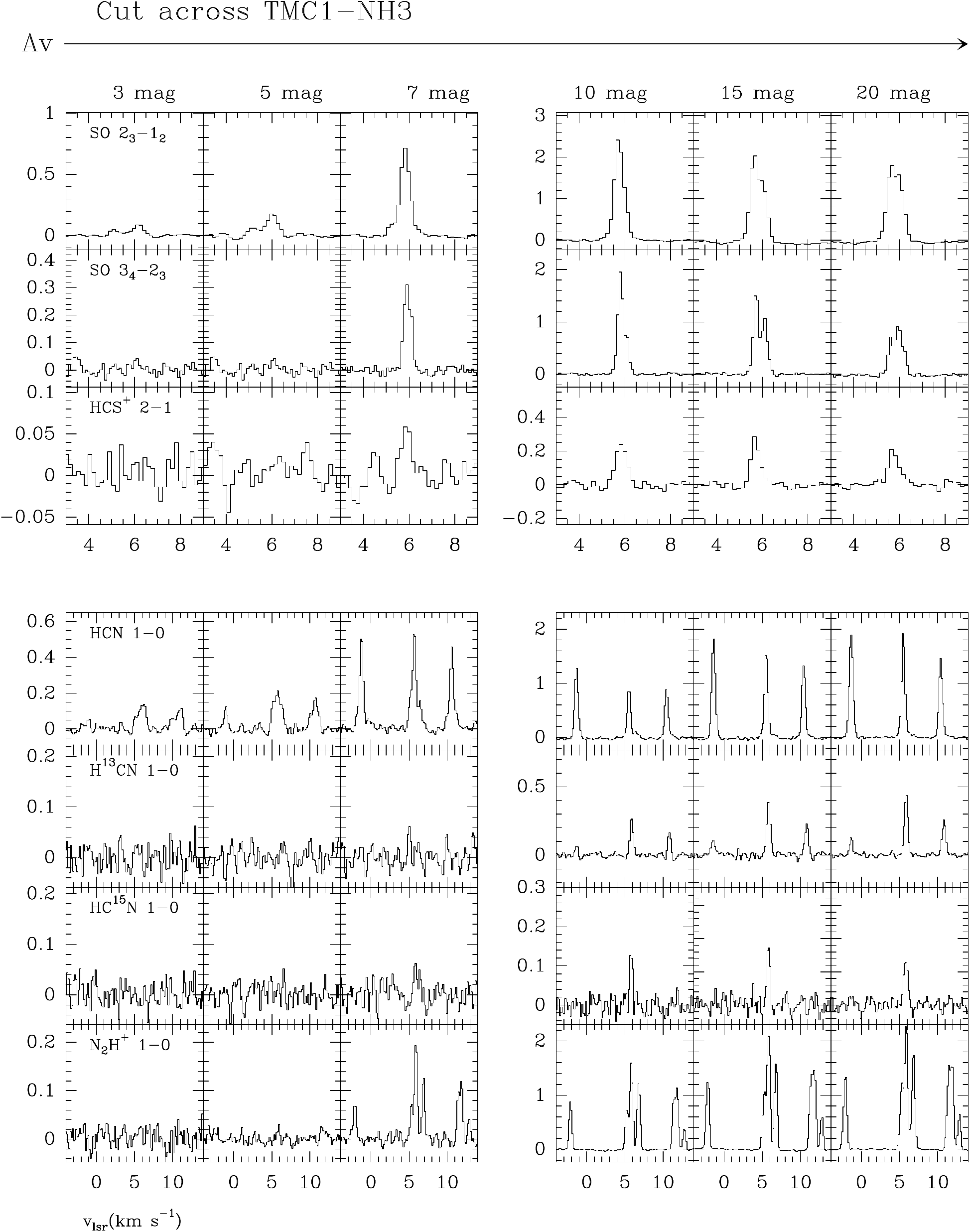}
\caption{The same as Fig. B.3}
\label{}
\vspace{-0.1cm}
\end{figure*}

\begin{figure*}
\includegraphics[angle=0,scale=.9]{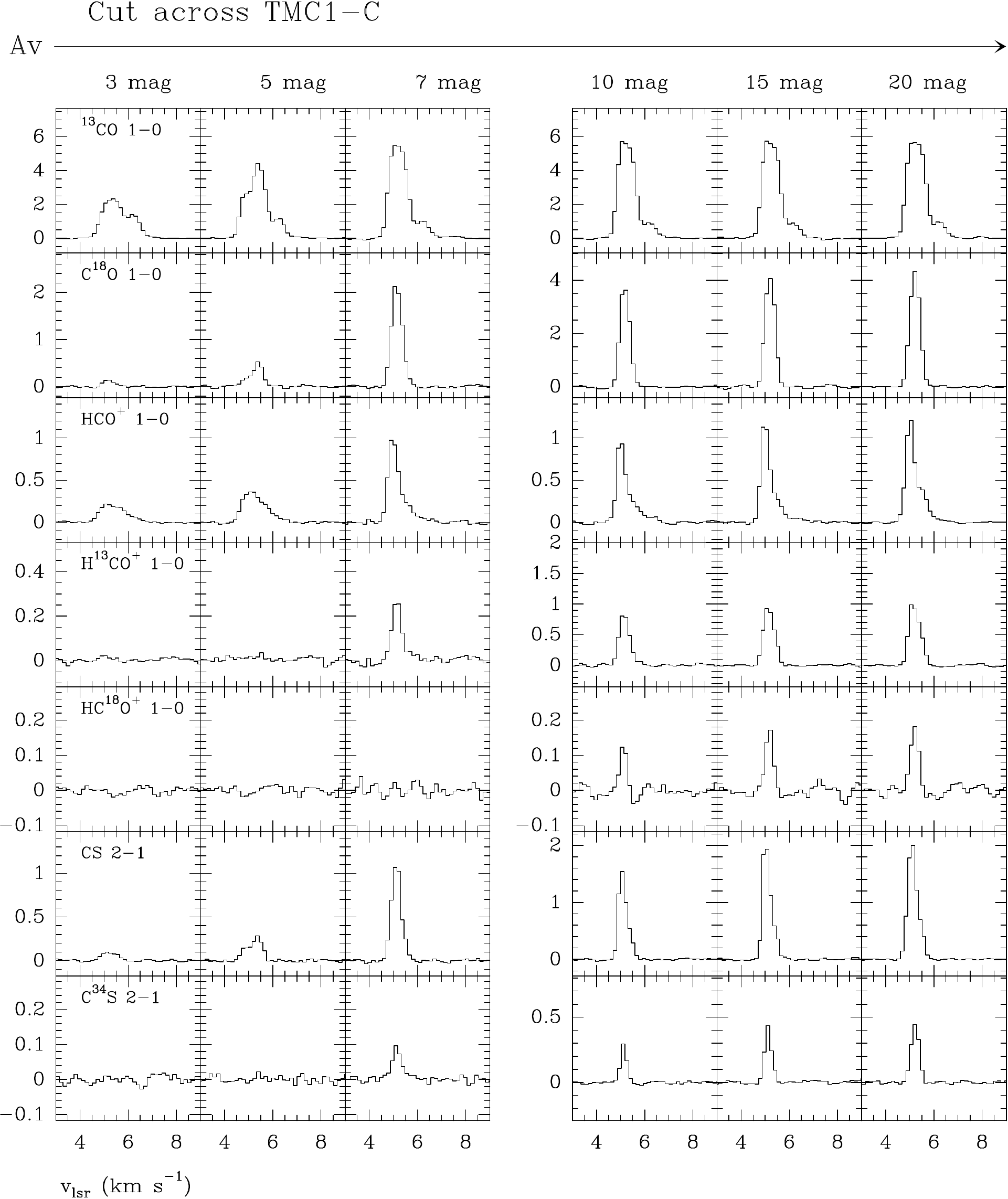}
\caption{Selected sample of spectra as observed with the 30m telescope towards the TMC~1-C cut (in T$_{MB}$).}
\label{c2-espec1}
\vspace{-0.1cm}
\end{figure*}

\begin{figure*}
\includegraphics[angle=0,scale=.9]{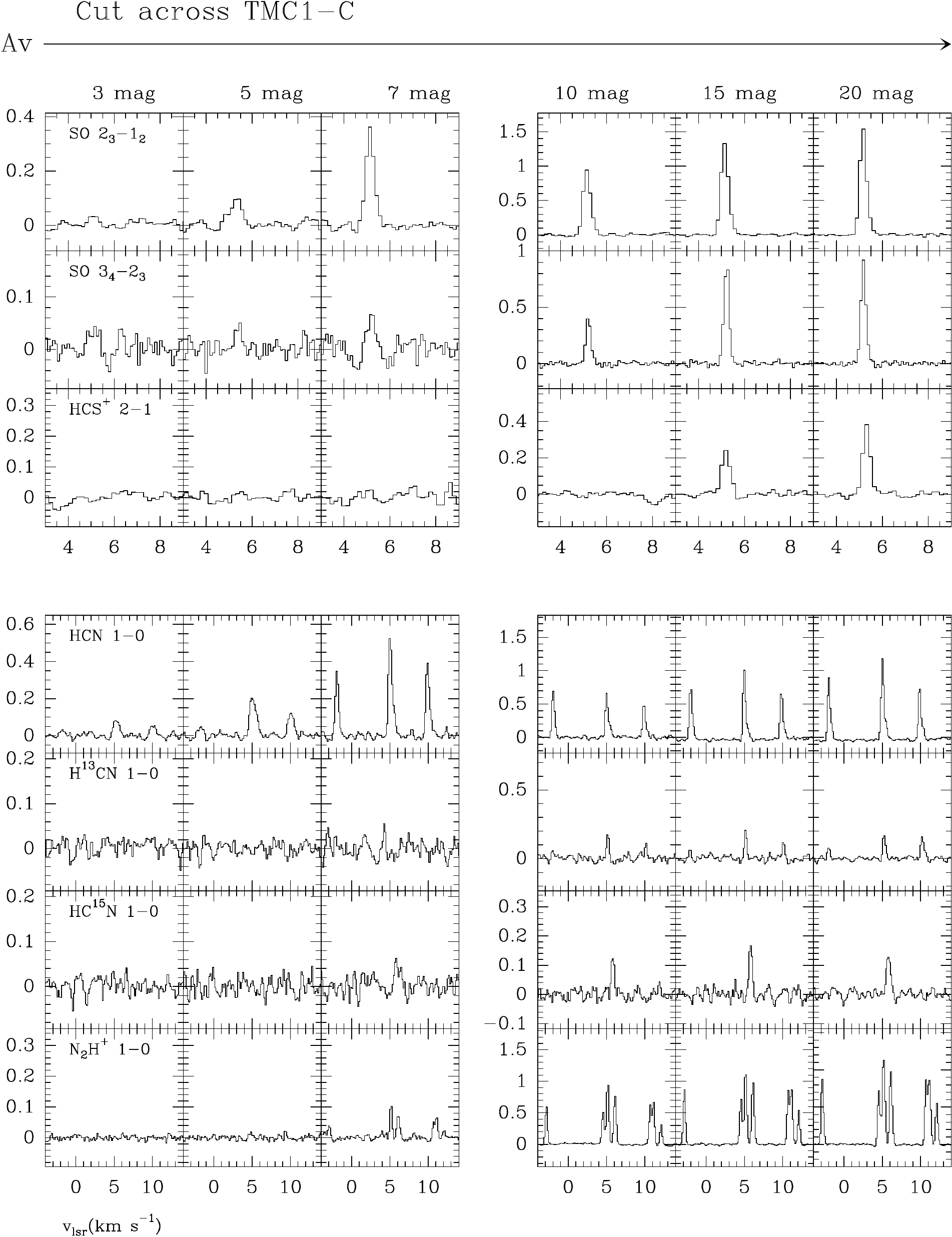}
\caption{The same as Fig. B.5}
\label{c2-espec2}
\vspace{-0.1cm}
\end{figure*}

\end{document}